\documentclass[11pt,a4paper,useAMS,usenatbib]{emulateapj}
\usepackage{epsfig}
\usepackage{amsmath}
\usepackage{natbib}

\begin{document}

\title{The diverse hot gas content and dynamics \\ of optically similar low-mass elliptical galaxies}

\author{\'Akos Bogd\'an\altaffilmark{1}, Laurence P. David,  Christine Jones, William R. Forman, and Ralph P. Kraft}
\affil{Smithsonian Astrophysical Observatory, 60 Garden Street, Cambridge, MA 02138, USA}
\email{E-mail: abogdan@cfa.harvard.edu \\  $^1$Einstein Fellow}

\shorttitle{HOT GAS CONTENT OF LOW-MASS ELLIPTICAL GALAXIES}
\shortauthors{BOGD\'AN ET AL.}

\begin{abstract}
The presence of hot X-ray emitting gas is ubiquitous in massive early-type galaxies. However, much less is known about the content and physical status of the hot X-ray gas in low-mass ellipticals. In the present paper we study the X-ray gas content of four low-mass elliptical galaxies using archival \textit{Chandra} X-ray observations. The sample galaxies, NGC821, NGC3379, NGC4278, and NGC4697, have approximately identical K-band luminosities, and hence stellar masses, yet their X-ray appearance is strikingly different. We conclude that the unresolved emission in NGC821 and NGC3379 is built up from a multitude of faint compact objects, such as coronally active binaries and cataclysmic variables. Despite the non-detection of  X-ray gas, these galaxies may host low density, and hence low luminosity, X-ray gas components, which undergo a Type Ia supernova (SN Ia) driven outflow. We detect hot X-ray gas with a temperature of $kT\sim0.35$ keV in NGC4278, the component of which has a steeper surface brightness distribution than the stellar light. Within the central $50\arcsec$ ($\sim$$3.9$ kpc) the estimated gas mass is $\sim$$3\times10^{7} \ \rm{M_{\odot}}$, implying a gas mass fraction of $\sim$$0.06\%$.  We demonstrate that the X-ray gas exhibits a bipolar morphology in the northeast-southwest direction, indicating that it may be outflowing from the galaxy. The mass and energy budget of the outflow can be maintained by evolved stars and SNe Ia, respectively. The X-ray gas in NGC4697 has an average temperature of $kT\sim0.3$ keV, and a significantly broader distribution than the stellar light. The total gas mass within $90\arcsec$ ($\sim$$5.1$ kpc) is  $\sim$$2.1\times10^{8} \ \rm{M_{\odot}}$, hence the gas mass fraction is $\sim$$0.4\%$. Based on the distribution and physical parameters of the  X-ray gas, we conclude that it is most likely in hydrostatic equilibrium, although a subsonic outflow may be present.

\end{abstract}


\section{Introduction}
Although the hot X-ray emitting gas content of massive ($M_{\star}>10^{11} \ \rm{M_{\odot}}$) elliptical galaxies has been studied in  detail \citep[e.g.][]{forman85,mathews03,randall06,forman07,kraft11}, the X-ray gas content of low-mass ellipticals is less explored. The major issue in studying low-mass ellipticals ($M<10^{11} \ \rm{M_{\odot}}$) is their relatively X-ray faint nature. In particular, the observed X-ray emission from low-mass elliptical galaxies is dominated by the populations of resolved and unresolved compact X-ray sources. Besides the populations of compact objects, emission from hot X-ray gas may also be present, the detection of which may be compromised by the underlying emission from unresolved compact objects. Moreover, given the shallow potential well of such galaxies, the hot X-ray emitting gas may not be in a stationary state. SNe Ia might be energetically capable of driving galactic-scale outflows \citep{david06}. In agreement with this result, a galactic-scale outflow had been detected in the bulge of M31 \citep{li07,bogdan08} and in the Sombrero galaxy \citep{li11}.

\begin{figure*}[t]
  \begin{center}
    \leavevmode
      \epsfxsize=8.5cm\epsfbox{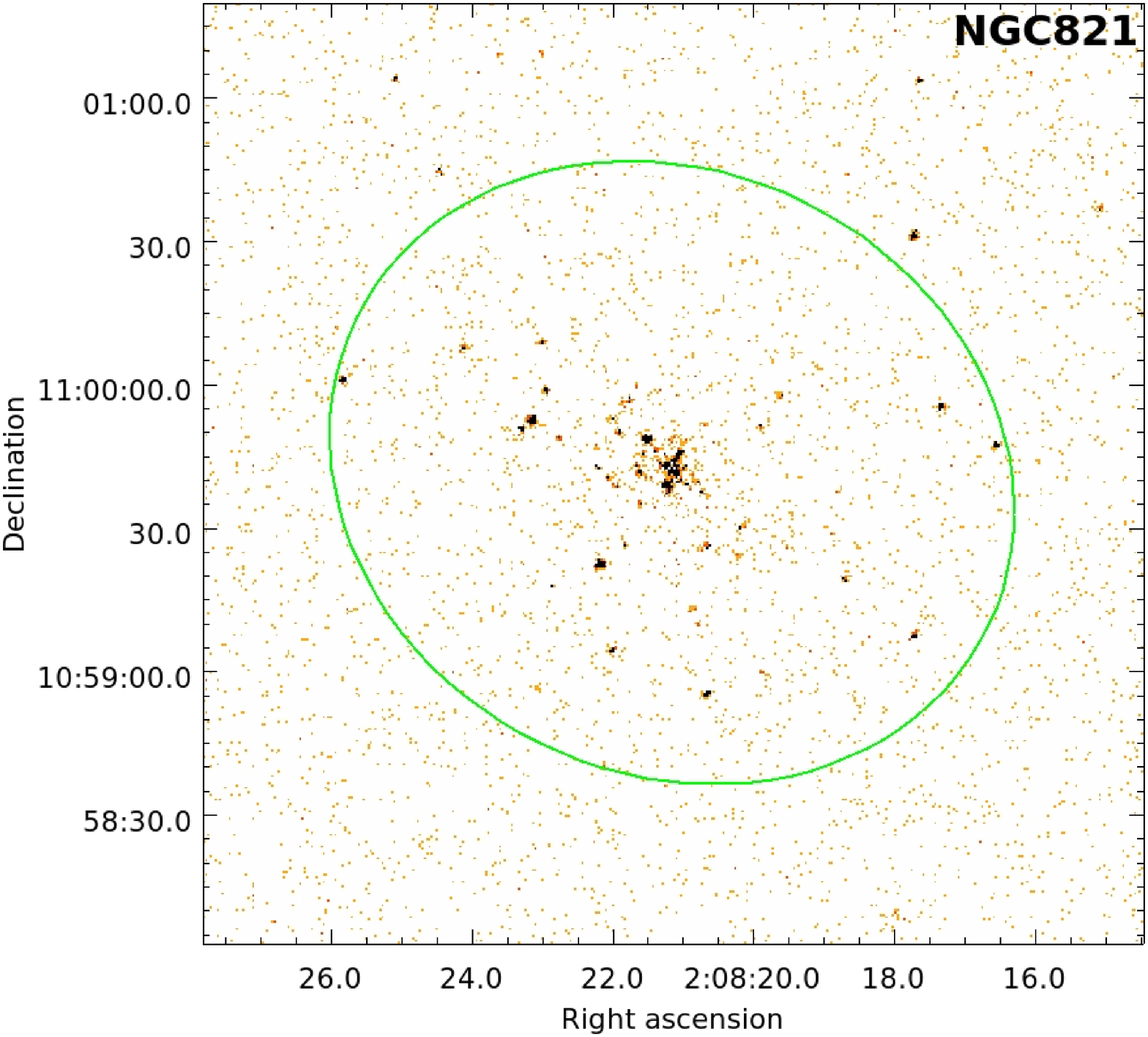}
\hspace{0.4cm} 
      \epsfxsize=8.5cm\epsfbox{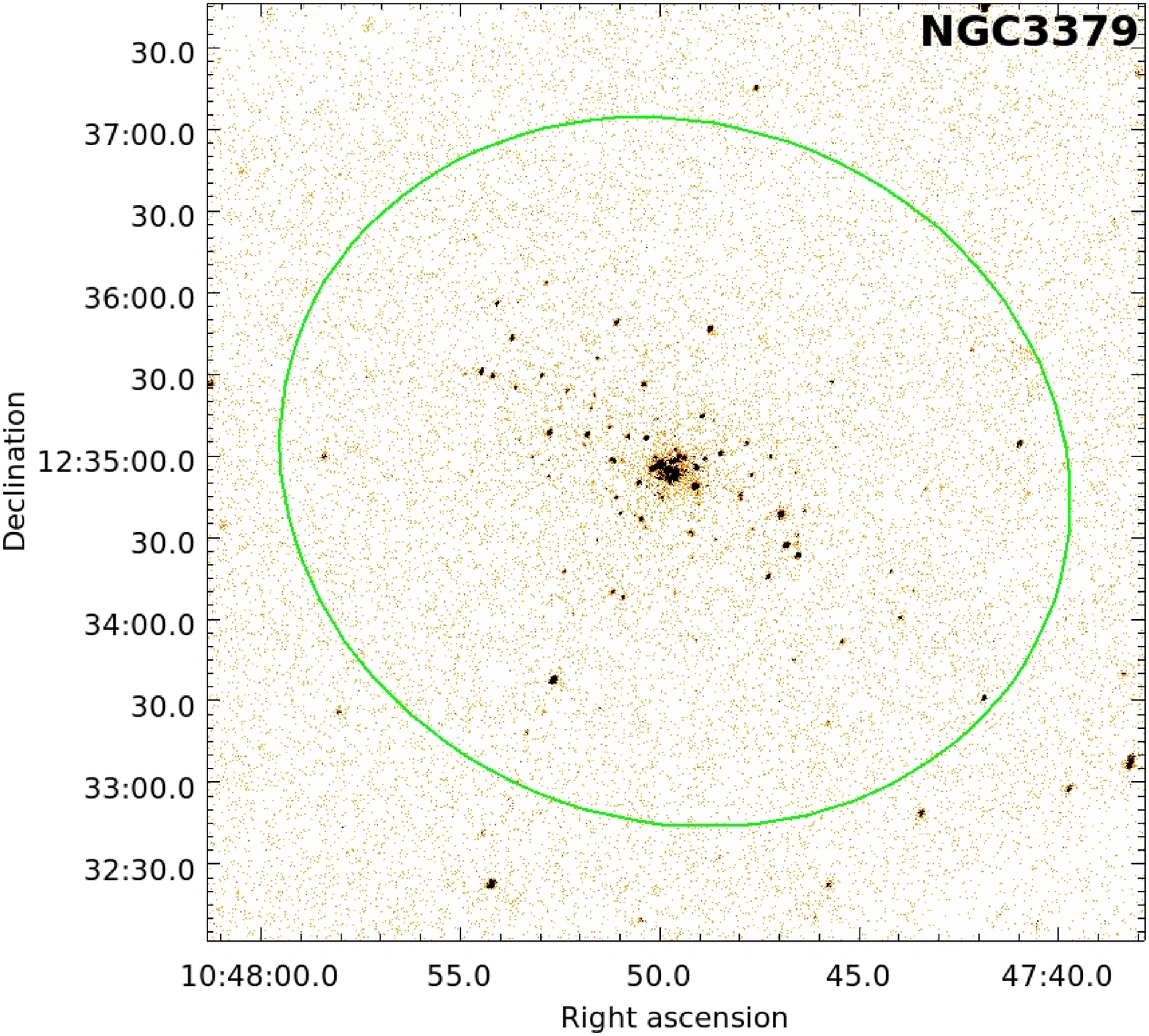}
     \epsfxsize=8.5cm\epsfbox{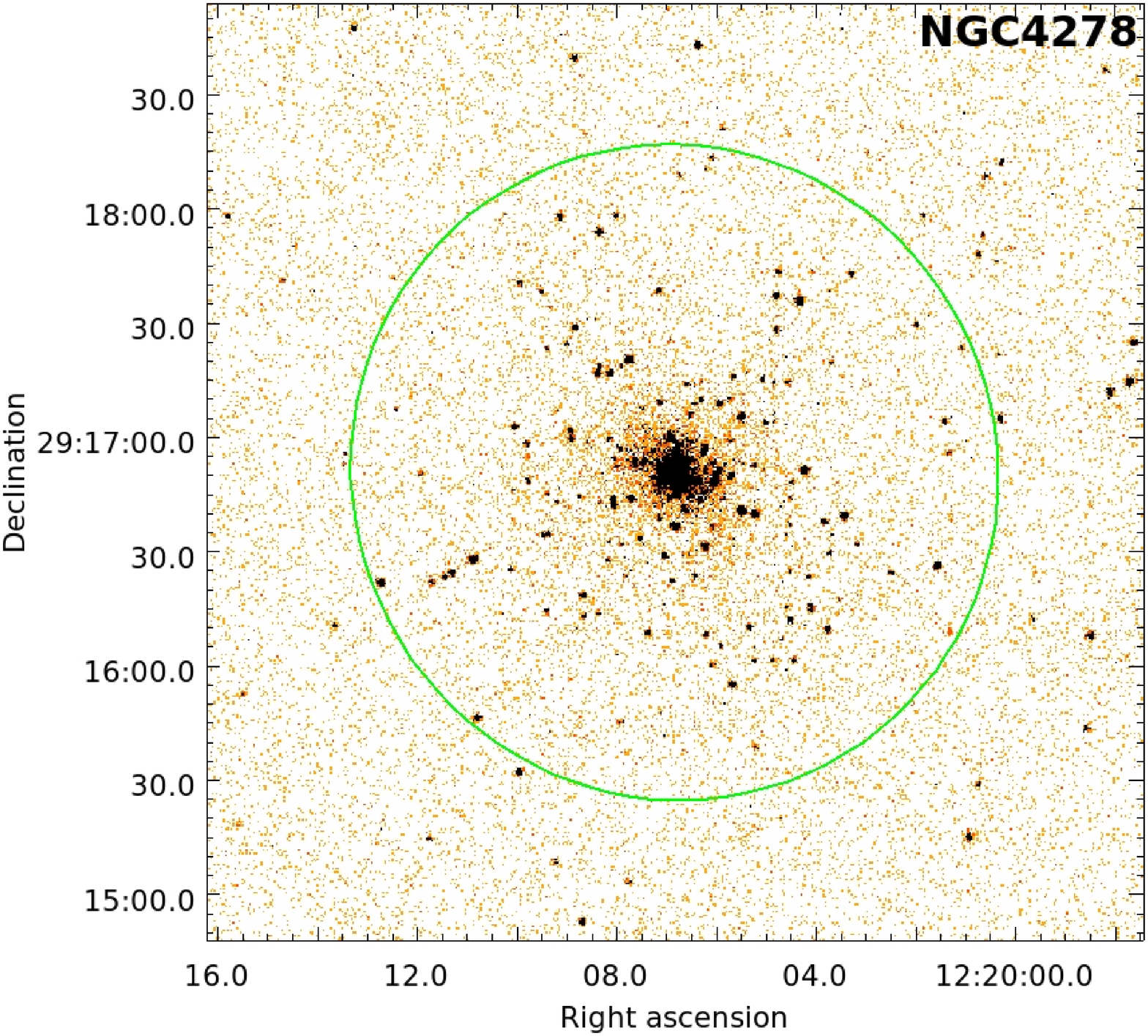}
\hspace{0.4cm} 
      \epsfxsize=8.5cm\epsfbox{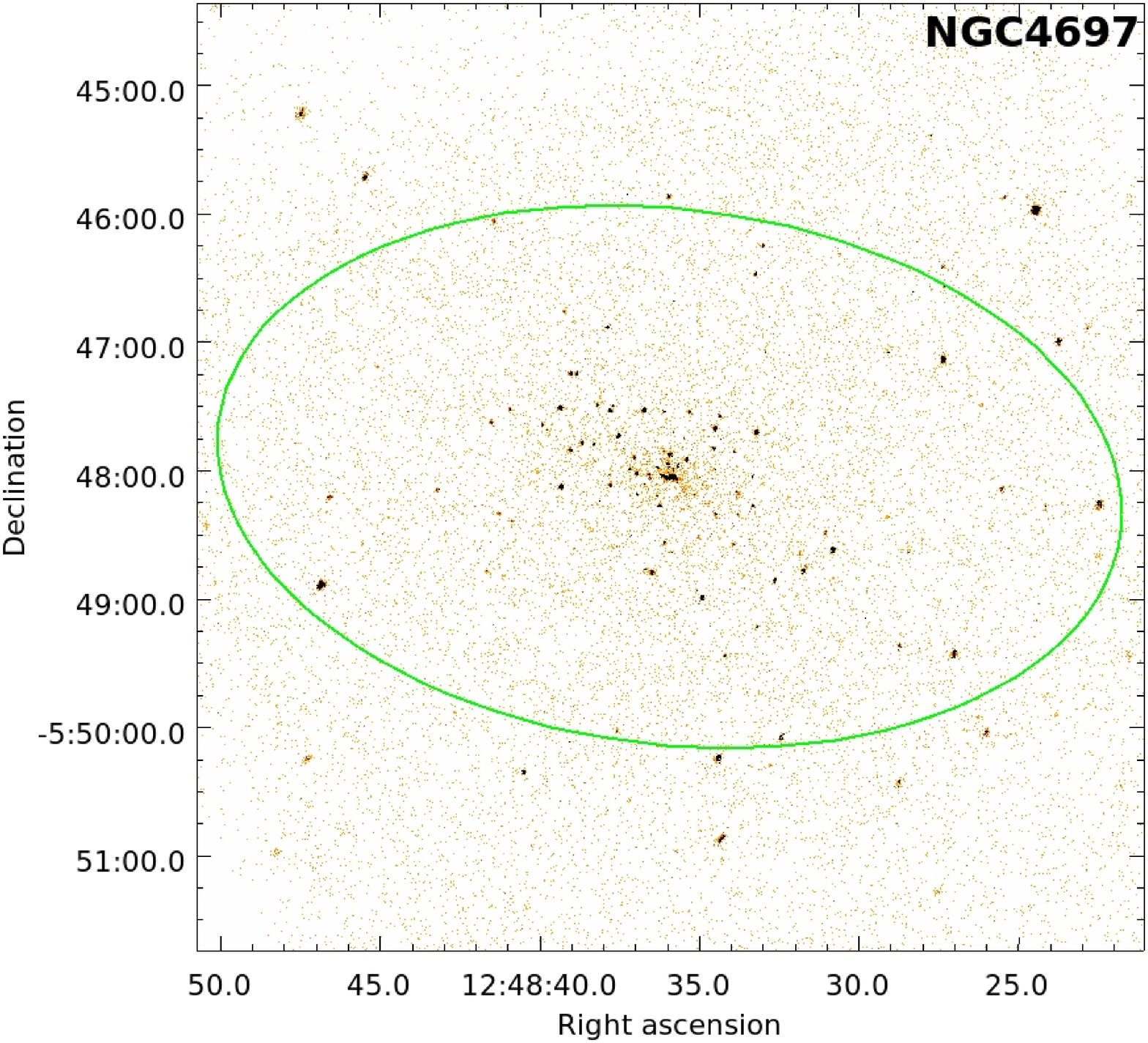}
      \caption{$0.5-2$ keV band raw \textit{Chandra} images of the four analyzed galaxies. The majority of the bright point sources are low-mass X-ray binaries associated with the galaxies, while minor fraction of them are background AGN. The overplotted elliptic regions show the $D_{\rm{25}}$ ellipses, whose extent listed in Table \ref{tab:list2}.}
\vspace{0.5cm}
     \label{fig:img}
  \end{center}
\end{figure*}

Detailed X-ray studies of the Milky Way and nearby galaxies have led to a better understanding of the origin of the resolved and the unresolved X-ray emission. With the superior angular resolution of \textit{Chandra}, the population of bright low-mass X-ray binaries (LMXBs) could be studied in detail \citep[e.g.][]{gilfanov04}. Moreover, it has been established that (at least) a portion of unresolved X-ray emission is associated with the stellar population and is the superposition of a multitude of faint compact sources, such as active binaries (ABs) and cataclysmic variables  \citep[CVs;][]{sazonov06,revnivtsev06,revnivtsev09}. Thus, if sufficiently deep \textit{Chandra} observations are available, then the various X-ray emitting components can be identified in nearby galaxies, hence the diffuse emission from hot X-ray gas can be separated from the population of resolved and faint unresolved compact objects. This extensive knowledge permits us to determine whether low-mass ellipticals host X-ray emitting gas and, additionally, to address the state of the hot gas. 

In the present paper, we study four elliptical galaxies, whose optical appearances are nearly identical, yet their X-ray properties are strikingly different. The four galaxies, NGC821, NGC3379, NGC4278, and NGC4697, are relatively nearby ($D=9.8-24.1$ Mpc) and have deep \textit{Chandra} observations (Figure \ref{fig:img}). Therefore, we are able to identify and remove the bulk of the X-ray emission from LMXBs, hence bright point sources do not significantly contaminate the  diffuse emission. Furthermore, the low and stable instrumental background of \textit{Chandra} permits us to reliably study low surface brightness emission. The goals of this paper are twofold. First, we aim to unveil whether the sample galaxies host a significant amount of hot X-ray emitting gas. Second, if they do possess X-ray gas, we intend to measure its  properties, morphology, and understand its physical state.

\begin{table*}
\caption{Elliptical galaxies studied in this paper.}
\begin{minipage}{18.5cm}
\renewcommand{\arraystretch}{1.3}
\centering
\begin{tabular}{c c c c c c c c c c c}
\hline 
Name & Distance &  $L_{K} $    &  $M_{\star}/L_{K} $            &  $M_{\star}$ & $N_{H}$ & Morph.    & $ T_{\mathrm{obs}} $ & $ T_{\mathrm{filt}} $ & $L_{\mathrm{lim}}$            & $D_{\rm{25}}$ \\ 
     & (Mpc)    &($\mathrm{L_{K,\odot}}$) & ($M_{\odot}/L_{K,\odot} $) & ($M_{\odot}$) & (cm$^{-2}$) & type         & (ks)  & (ks)                  &  ($ \mathrm{erg \ s^{-1}} $)  & ($2a, 2b,  \theta $)    \\ 
     &   (1)    &            (2)           &   (3)      &     (4)     &      (5)                 &   (6) &  (7)                  &   (8)                         & (9)  & (10)  \\
\hline 
NGC821         & $ 24.1^a  $ & $ 8.2 \times 10^{10}$ & 0.83 & $ 6.8 \times 10^{10}$  & $  6.4 \times 10^{20}$ &  E6  & $ 229.7 $  & $  187.6 $& $ 2 \times10^{37} $ & $ 2.45\arcmin, 2.10 \arcmin, 26.0\degr   $ \\
NGC3379        & $ 9.8^b   $ & $  6.1 \times 10^{10}$& 0.83 &  $ 5.1 \times 10^{10}$  &$  2.9 \times 10^{20}$ &  E1  &$ 341.4 $ & $ 310.0 $& $ 2 \times 10^{36} $ & $ 4.90\arcmin, 4.27 \arcmin, 71.0\degr   $  \\
NGC4278     & $ 16.1^a  $ & $  7.1 \times 10^{10}$& 0.82 &$ 5.8 \times 10^{10}$  & $  1.8 \times 10^{20}$ &  E1-2 & $  476.7 $ & $  437.5 $& $ 4 \times 10^{36} $ & $ 2.88\arcmin, 2.82 \arcmin, 27.5\degr   $  \\ 
NGC4697     & $ 11.8^a  $ & $  8.1 \times 10^{10}$& 0.82 &$ 6.6 \times 10^{10}$  & $  2.1 \times 10^{20}$ &  E6           &  $ 195.7 $ & $ 159.9 $& $ 5 \times 10^{36} $ & $ 7.08\arcmin, 4.17 \arcmin, 83.1\degr   $  \\
\hline \\
\end{tabular} 
\end{minipage}
\textit{Note.} Columns are as follows. (1) References are: $^a$ \citet{tonry01}  -- $^b $ \citet{m105distance}. (2) Total K-band luminosity. (3) K-band mass-to-light ratios computed from \citet{bell03} using the $B-V$ color indices of galaxies \citep{devaucouleurs91}. (4) Total stellar mass based on the K-band luminosity and the K-band mass-to-light ratios. (5) Galactic absorption \citep{dickey90}. (6) Morphological type, taken from NED (http://nedwww.ipac.caltech.edu/). (7) and (8) \textit{Chandra} exposure times before and after flare filtering. (9) Source detection sensitivity in the $ 0.5-8 $ keV energy range. (10) Major axis diameter, minor axis diameter, and position angle of the $D_{\rm{25}}$ ellipse. \\

\label{tab:list2}
\end{table*}  

Since the X-ray emitting components of NGC821, NGC3379, NGC4278, and NGC4697 have already been analyzed to a certain extent by other authors, we briefly review previous works. The resolved and unresolved X-ray emitting components of NGC821 have been studied by \citet{pellegrini07} using \textit{Chandra} X-ray observations. In addition to their thorough analysis of bright LMXBs, they also concluded that the bulk of the unresolved emission originates from unresolved LMXBs and that hot X-ray gas may only be present in the inner $10\arcsec$ region. The unresolved X-ray emission of NGC3379 was first investigated by \citet{david05}, based on a $32$ ks \textit{Chandra} observation. They concluded that X-ray gas may be present within the central $15\arcsec$ region with a temperature of $kT=0.6$ keV. Based on a much deeper, $341$ ks, \textit{Chandra} data set, \citet{revnivtsev08}  claimed that NGC3379 is virtually gas free, and the unresolved X-ray emission is built up from faint undetected compact objects, mostly by ABs and CVs. However, based on the same data set, \citet{trinchieri08} reported that NGC3379 hosts hot X-ray gas in the central $15\arcsec$ region, which is outflowing from the galaxy. Although the population of resolved sources in NGC4278 has been extensively studied \citep[e.g.][]{kim06}, the unresolved emission from NGC4278 has not been investigated in detail. The unresolved X-ray emitting components of NGC4697 were explored by \citet{sarazin01} based on a $40$ ks \textit{Chandra} observation. They demonstrated the presence of X-ray gas, which has a significantly broader distribution than the stellar light. However, more recently NGC4697 was observed by \textit{Chandra} for an additional $\sim$$150$ ks, allowing a better separation of bright LMXBs and  truly diffuse emission. Additionally, accurate calibration of the population of faint compact objects only became available relatively recently. Hence, in some of the previous works, the emission from these sources could not be precisely disentangled from the hot X-ray gas. Therefore, our uniform analysis of the sample galaxies with the most up-to-date calibration of LMXBs and faint compact objects can lead to a better understanding of the X-ray gas content of low-mass ellipticals.

The paper is structured as follows. In Section 2, we introduce the analyzed galaxy sample. In Section 3, we describe the reduction of the data. In Section 4, we study the X-ray gas content of the sample galaxies. The physical state of the detected  X-ray gas is discussed in Section 5, and we summarize our results in Section 6.

\section{Sample selection}
To identify the most suitable galaxies for our analysis, we rely on the volume limited $\mathrm{ATLAS^{3D}}$ catalog, which  consists of 260 nearby early-type galaxies \citep{cappellari11}. To study low-mass ellipticals that may host a notable amount of hot X-ray gas, we selected galaxies with $M_K > -24 $ mag ($L_K<8.2\times10^{10} \ \rm{L_{K,\odot}}$) and with morphological types  $T<-1.5$. Finally, we filtered galaxies by distance and selected those with distances $D<30$ Mpc. The  resulting list includes a sample of 41 galaxies. 

In order to study the unresolved emission and unveil the presence of a possible hot X-ray gas component, the bulk of  the luminosity from LMXBs must be removed. Therefore, we demand a minimum source detection sensitivity of $\sim$$3\times10^{37} \ \rm{erg \ s^{-1}}$, which assures that at least $\sim$$75\%$ of the luminosity from LMXBs can be subtracted \citep{gilfanov04}. To identify galaxies with sufficiently deep \textit{Chandra} exposures, the list of 41 galaxies from the $\mathrm{ATLAS^{3D}}$ catalog was cross-correlated with the \textit{Chandra} archive. We find that four nearby low-mass elliptical galaxies have deep enough \textit{Chandra} data to resolve sources with luminosities exceeding $3\times10^{37} \ \rm{erg \ s^{-1}}$. These four galaxies are  NGC821, NGC3379, NGC4278, and NGC4697. Basic properties of the sample galaxies are listed in Table \ref{tab:list2}. 

The remaining sample of $37$ low-mass elliptical galaxies, identified in the $\mathrm{ATLAS^{3D}}$ catalog, have source detection sensitivities of $>$$3\times10^{37} \ \rm{erg \ s^{-1}}$. Their hot X-ray gas content and the physical state of the gas will be discussed in a forthcoming study (L. David et al. in preparation).

\section{Data reduction}
\subsection{\textit{Chandra}}
\label{sec:chandra}
To study the various X-ray emitting components of the sample galaxies, we rely on \textit{Chandra} observations. The list of analyzed observations is given in Table \ref{tab:list1}. To reduce the data, we use the standard \textsc{CIAO} software package tools\footnote{http://cxc.harvard.edu/ciao/} (\textsc{CIAO} version 4.3; \textsc{CALDB} version 4.4.2).

As a first step in the data preparation, we filter the flare contaminated time intervals for each observation, following the method described in \citet{bogdan08}. On average, the clean exposure times are $\sim$$10\%-15\%$ shorter than the original exposures (Table \ref{tab:list1}). For each galaxy, the flare-filtered observations were combined and merged to the coordinate system of the observation with the longest exposure time.

To detect point sources, we use  the \textsc{CIAO} \textsc{wavdetect} tool on the merged unfiltered data set in the $0.5-8$ keV energy range. As our particular aim is to study the diffuse emission, we changed the values of several parameters to obtain larger source cells -- for a detailed description see \citet{bogdan08}. Since NGC3379 and NGC4278 host ultraluminous X-ray sources at their centers, spill-over counts from these sources could add a notable contribution to the unresolved emission. To avoid this problem, we excluded the central $7\arcsec$ and $5\arcsec$ radius regions of NGC3379 and NGC4278. We note that for point source detection, we used the original unfiltered data set, since the longer exposures outweigh the high background periods, thereby resulting in a better source detection sensitivity. The lists of detected sources were used to  mask out the point sources for further study of the diffuse emission. To correct for vignetting and to estimate source detection sensitivities, exposure maps were produced using a power law model with a slope of  $\Gamma=1.56$, typical for LMXBs \citep{irwin03}. Assuming this spectrum, we also estimated the source detection sensitivity of the combined observations, which for all four galaxies is better than $\sim$$3\times10^{37} \ \rm{erg \ s^{-1}}$. 

To subtract the background, we use nearby regions on the ACIS-S3 CCD for all galaxies but NGC4697. This can be done since the optical extent of the galaxies does not fill the field-of-view of the ACIS-S3 CCD. For  NGC4697, we use extra care in subtracting the background components for two reasons. First, it is located in a galaxy group, and second,  the galaxy is located in the north polar spur. Therefore, we use the ACIS-S1 CCD to determine the total background component, which is composed of the instrumental background components and the sky background, which includes the soft Galactic emission, the population of unresolved cosmic X-ray background sources, and the emission associated with the north polar spur. 

\begin{table}
\caption{The list of analyzed \textit{Chandra} observations.}
\begin{minipage}{8.75cm}
\renewcommand{\arraystretch}{1.3}
\centering
\begin{tabular}{c c c c c }
\hline 
Galaxy & Obs ID &  $T_{\rm{obs}}$ (ks) & $T_{\rm{filt}}$ (ks) & Instrument \\
\hline
NGC821 & 4006 & 13.7 & 9.5 & ACIS-S \\
NGC821 & 4408 & 25.3 & 5.5 & ACIS-S \\
NGC821 & 5691 & 40.1 & 35.9 & ACIS-S\\
NGC821 & 5692 & 28.0 & 24.6 & ACIS-S\\
NGC821 & 6310 & 32.4 & 29.3 & ACIS-S\\
NGC821 & 6313$^{\dagger}$ & 50.1 & 46.7 & ACIS-S\\
NGC821 & 6314 & 40.1 & 36.1 & ACIS-S\\

NGC3379 & 1587 & 31.9 & 24.8 & ACIS-S \\
NGC3379 & 7073$^{\dagger}$ & 85.2 & 75.8 & ACIS-S \\
NGC3379 & 7074 & 70.0 & 65.5 & ACIS-S \\
NGC3379 & 7075 & 84.2 & 78.0 & ACIS-S \\
NGC3379 & 7076 & 70.1 & 65.9 & ACIS-S \\

NGC4278 & 4741 & 37.9 & 34.3 & ACIS-S \\
NGC4278 & 7077 & 111.7 & 101.9 & ACIS-S \\
NGC4278 & 7078 & 52.1 & 45.8 & ACIS-S \\
NGC4278 & 7079 & 106.4 & 99.0 & ACIS-S \\
NGC4278 & 7080 & 56.5 & 52.9 & ACIS-S \\
NGC4278 & 7081$^{\dagger}$  & 112.1 & 103.6 & ACIS-S \\

NGC4697 & 784   & 39.8 & 37.3 & ACIS-S \\
NGC4697 & 4727 & 40.5 & 36.0 & ACIS-S \\
NGC4697 & 4728 & 36.2 & 30.9 & ACIS-S \\
NGC4697 & 4729 & 38.6 & 19.2 & ACIS-S \\
NGC4697 & 4730$^{\dagger}$  & 40.6 & 36.5 & ACIS-S \\

\hline \\
\end{tabular} 
\end{minipage}
$^{\dagger}$ The coordinate system of these observations  was used as reference when merging observations.   \\

\label{tab:list1}
\end{table}

\subsection{Two-Micron All Sky Survey}
To derive the stellar mass and trace the stellar light of the sample galaxies, we rely on the K-band data of the Two-Micron All Sky Survey (2MASS) Large Galaxy Atlas  \citep{jarrett03}. The K-band images, provided by the 2MASS archive, are background subtracted for all galaxies, except for  NGC4278. Therefore, we used regions nearby NGC4278 to estimate and subtract the  background level. The observed background-subtracted K-band counts ($S$) were converted to physical units using
\begin{eqnarray}
m_{\rm{K}}= \rm{KMAGZP} - 2.5 \log S \ , 
\end{eqnarray}
where $m_{\rm{K}}$ is the apparent K-band magnitude, and $\rm{KMAGZP}$ is the zero-point magnitude given in the image header. For each galaxy we compute $m_{\rm{K}}$ using Equation (1), which we convert to absolute magnitude and further to luminosity, assuming that the absolute K-band magnitude of the Sun is $ M_{K,\odot} = 3.28 $ mag.

The total K-band luminosities of the sample galaxies are very similar, and lie in the range of $(6.1-8.2)\times10^{10} \ \rm{L_{K,\odot}}$. Based on the K-band luminosity and the K-band mass-to-light ratios ($M_{\star}/L_K$) of the galaxies, we compute their total stellar masses. To compute the  $M_{\star}/L_K$ ratios, we rely on the $B-V$ color indices \citep{devaucouleurs91} and the results of galaxy evolution modeling \citep[][Table \ref{tab:list2}]{bell03}. Since the derived $M_{\star}/L_K$ ratios are very similar in each galaxy, the resulting stellar masses are also compatible and are in the range of  $(5.1-6.8)\times10^{10} \ \rm{M_{\odot}}$ (Table \ref{tab:list2}).

\begin{figure*}[t]
  \begin{center}
    \leavevmode
      \epsfxsize=8.5cm\epsfbox{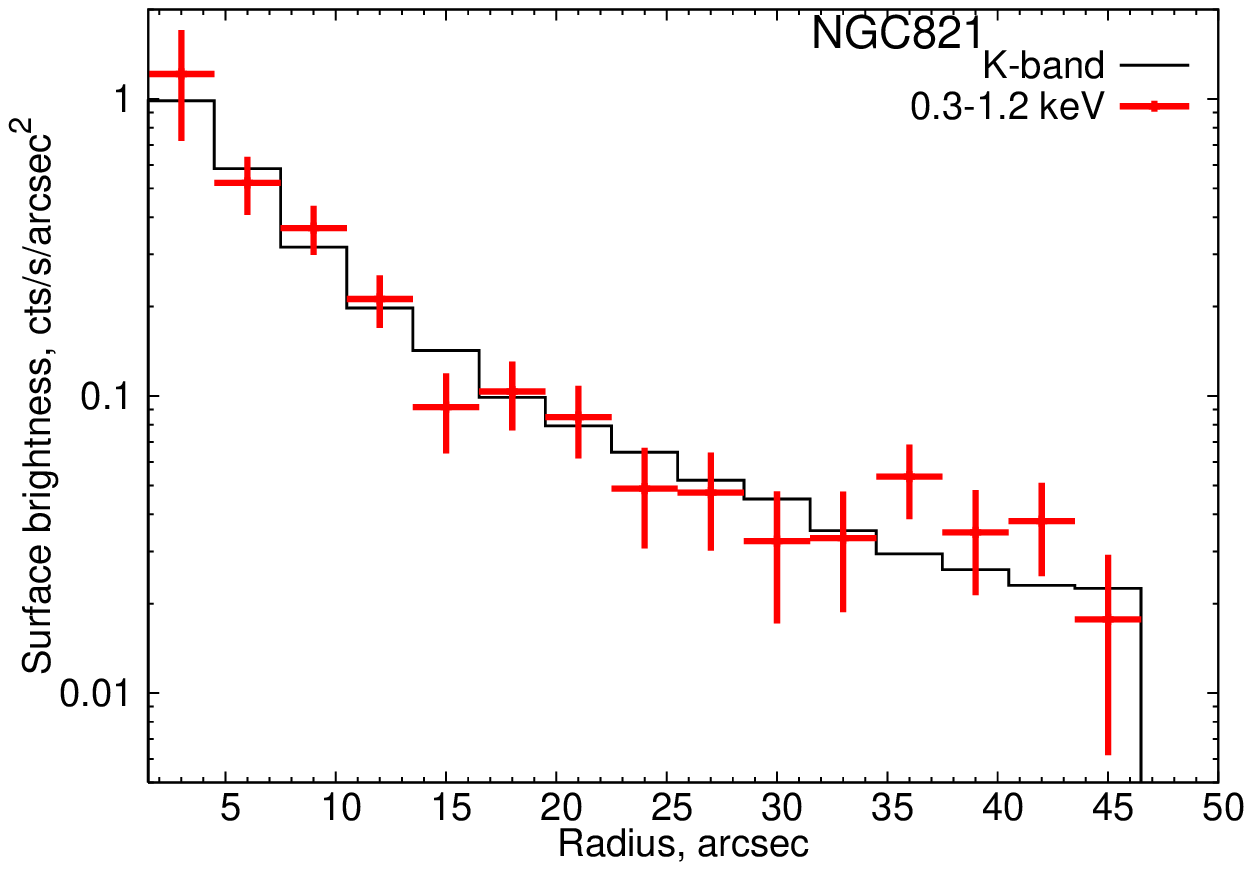}
\hspace{0.4cm} 
      \epsfxsize=8.5cm\epsfbox{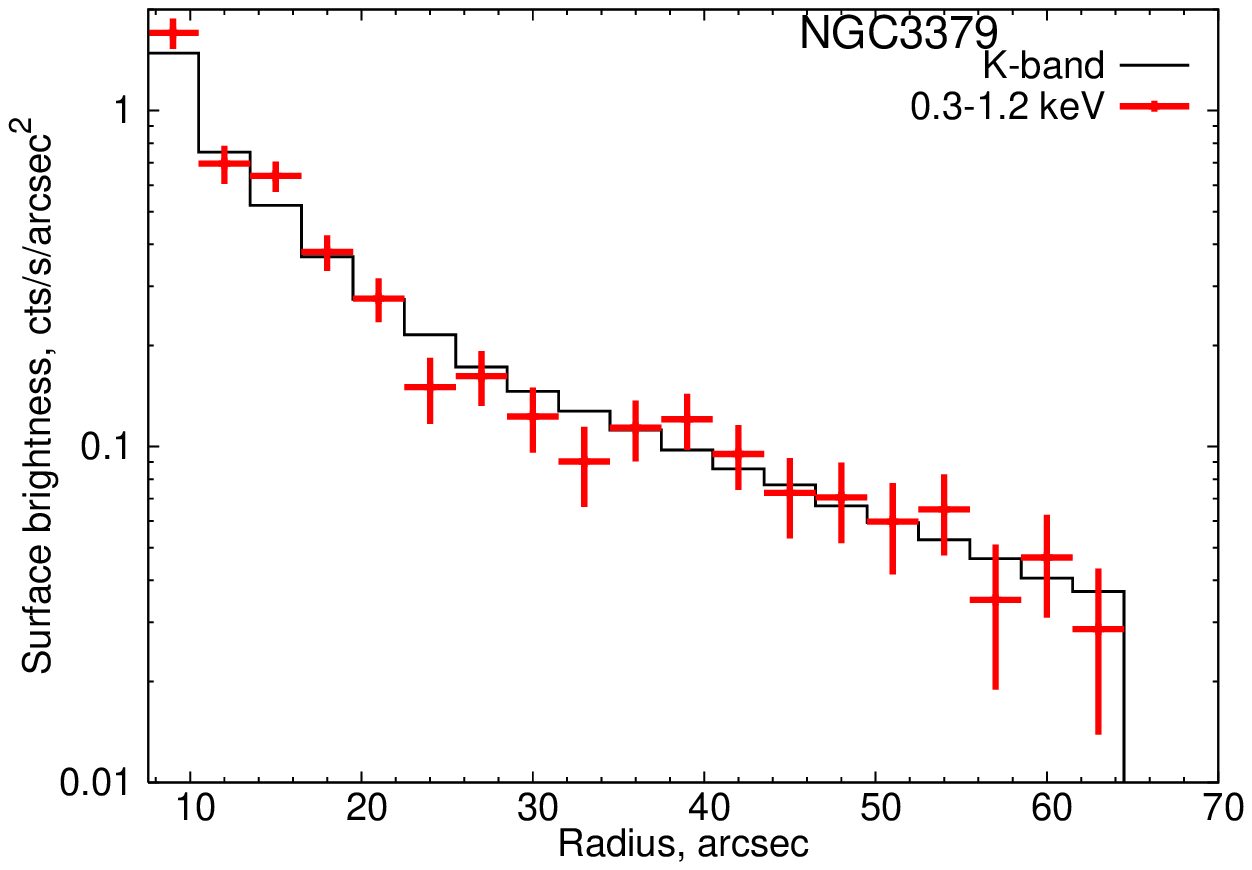}
     \epsfxsize=8.5cm\epsfbox{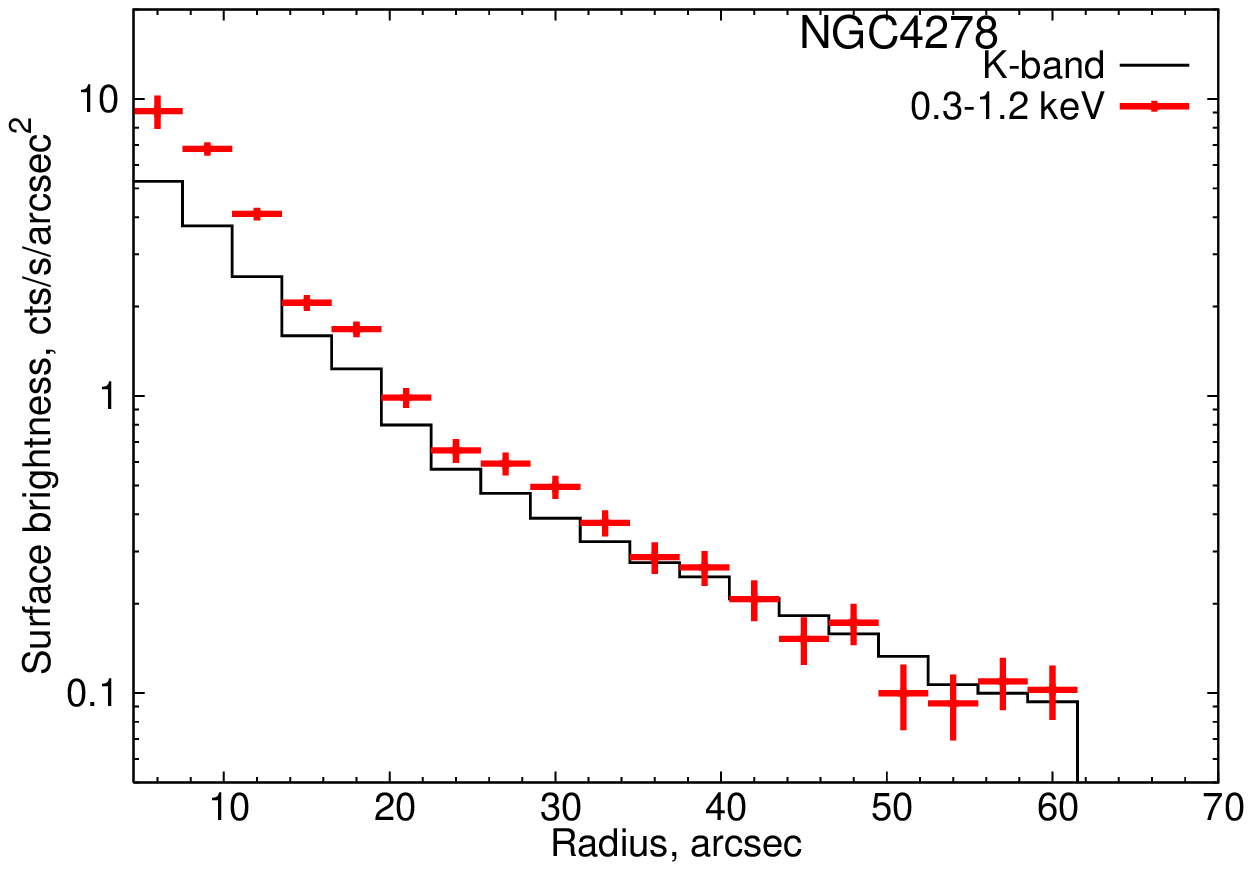}
\hspace{0.4cm} 
      \epsfxsize=8.5cm\epsfbox{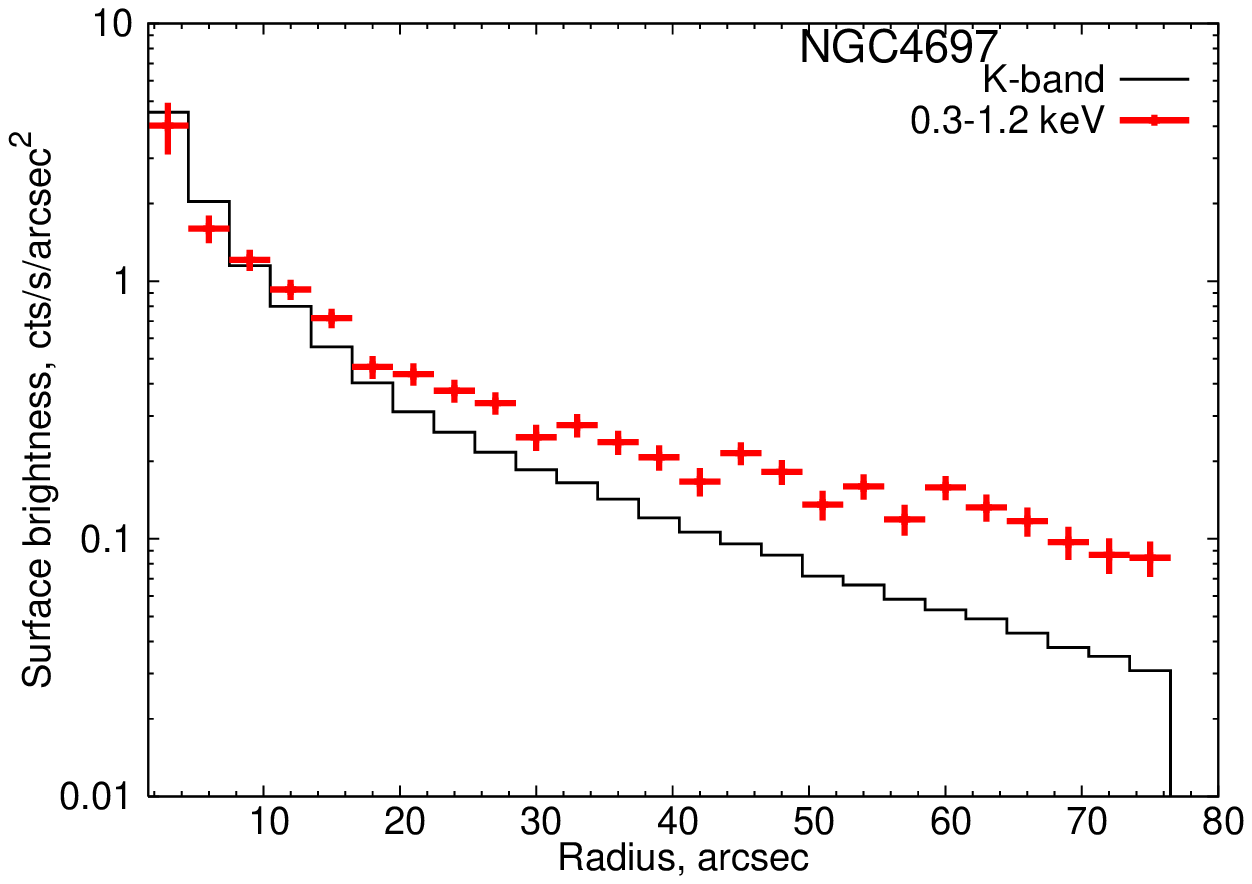}
      \caption{Surface brightness distribution of the $0.3-1.2 $ keV band unresolved X-ray emission for the sample galaxies. To construct the profiles, we used circular annuli centered on the centroid of each galaxy. Vignetting correction is applied and the background components are subtracted. The X-ray light distribution is compared with the K-band light profiles, which are obtained using the same regions.  The K-band profiles are  arbitrarily normalized.}
\vspace{0.5cm}
     \label{fig:profile}
  \end{center}
\end{figure*}

\begin{table*}
\caption{$L_X/L_K$ ratios observed for the sample galaxies}
\begin{minipage}{18cm}
\renewcommand{\arraystretch}{1.3}
\centering
\begin{tabular}{c c c c c}
\hline 
Galaxy & $L_{\rm{0.5-2keV}}$ & $L_{\rm{0.5-2keV,XB,sub}}$ & $L_K$ & $L_{\rm{0.5-2keV,XB,sub}}/L_K$\\
           & ($\rm{erg \ s^{-1}}$)   & ($\rm{erg \ s^{-1}}$)      &  ($L_{\rm{K,\odot}}) $ & ($\rm{erg \ s^{-1} \ L_{K,\odot}}^{-1}$) \\
           &      (1)    & (2)  & (3)   & (4) \\
\hline
NGC821   & $2.9\times10^{38}$  & $2.8\times10^{38}$ & $8.2\times10^{10}$  & $3.4\times10^{27}$ \\
NGC3379 & $2.4\times10^{38}$  & $2.2\times10^{38}$ & $6.1\times10^{10}$  & $3.6\times10^{27}$ \\
NGC4278 & $1.2\times10^{39}$  & $1.1\times10^{39}$ & $7.1\times10^{10}$  & $1.5\times10^{28}$ \\
NGC4697 & $2.7\times10^{39}$  & $2.6\times10^{39}$  & $8.1\times10^{10}$ & $3.2\times10^{28}$ \\
\hline
M32$^{\dagger}$ & -- & --  & -- & $3.5\times10^{27}$\\ 
Solar neighborhood$^{\ddagger}$  & -- & --  & -- &$7.7\times10^{27}$ \\
\hline \\
\end{tabular} 
\end{minipage}
\textit{Note.} Columns are as follows. (1) Total observed $0.5-2$  keV band X-ray luminosity within the $D_{\rm{25}}$ ellipse after the removal of all detected point sources. The contribution of CXB sources and the contribution of source counts falling outside the source cells have been subtracted.  (2) Total observed $0.5-2$  keV band X-ray luminosity within the $D_{\rm{25}}$ ellipse after the emission from unresolved LMXBs is subtracted. (3) K-band luminosity of galaxies. (4) Contamination subtracted $L_X/L_K$ ratios in the $0.5-2$ keV energy range, computed from Columns (2) and (3).  \\
$^{\dagger}$\citet{bogdan11a}; $^{\ddagger}$ \citet{sazonov06}
\label{tab:list3}
\end{table*}

\section{The X-ray content of the galaxies}
To determine whether the sample galaxies host a notable amount of hot X-ray gas, we use three different approaches, discussed  throughout this section. First, we compare the surface brightness distribution of the soft ($0.3-1.2$ keV) band unresolved X-ray emission with that of the K-band light. Since the distribution of the faint compact objects follows the stellar light, a deviation from the K-band profile may indicate the presence of hot gas. Second, we compute X-ray-to-K-band luminosity ($L_X/L_K$) ratios of the unresolved X-ray emission in the $0.5-2$ keV energy band  and compare these with values obtained for admittedly gas-free galaxies and the Milky Way. An $L_X/L_K$ ratio significantly exceeding that of gas-free galaxies will demonstrate the presence of excess X-ray emission, presumably originating from hot gas. Finally, we extract X-ray energy spectra of the sample galaxies. If hot X-ray gas is present, then its physical properties can be measured by spectral fitting.

\subsection{X-ray surface brightness profiles}
\label{sec:profiles}
To construct X-ray surface brightness distributions of the unresolved emission, we extracted profiles in the $0.3-1.2$ keV energy range using circular annuli centered on the centroid of each galaxy.  For each profile, detected sources were excluded, a vignetting correction was applied, and the background level was subtracted. The X-ray light distribution is compared with the K-band profiles. To  obtain the K-band profiles, the same regions were used and the same source regions were excluded. The K-band profiles are normalized to approximately match the level of X-ray emission.  Although the normalization of the K-band profiles is arbitrary, the particular choice for NGC4278 and NGC4697 was motivated by the spatial distribution of the diffuse X-ray emission, discussed in Section \ref{sec:morphology}. The resulting profiles are shown in Figure \ref{fig:profile}.

Figure \ref{fig:profile} demonstrates that in two galaxies,   NGC3379 and NGC821, the X-ray light is well traced by the stellar light distribution at all radii. The good agreement between the X-ray and K-band profiles hints that the unresolved X-ray emission in these galaxies has a stellar origin, that is, it originates from the multitude of faint compact objects. However, in NGC4278 and NGC4697, the distribution of the  X-ray light shows deviations from the near-infrared profiles. In particular, the unresolved X-ray emission in NGC4278 has a steeper distribution in the central regions than the stellar light, whereas at larger radii it follows the K-band profile. Thus, an excess X-ray emitting component is present within the central regions of NGC4278. In NGC4697, the X-ray light exhibits a significantly broader distribution than the stellar light. This suggests that the bulk of the unresolved soft band X-ray emission arises from the excess X-ray emitting component, which is present at all radii. The nature and properties of the excess X-ray emitting components in NGC4278 and NGC4697 are extensively discussed in the following sections.

\subsection{$L_X/L_K$ luminosity ratios}
In old stellar populations, the $L_X/L_K$ ratio for the population of faint compact objects is fairly uniform, and is in the range of $L_{\rm{0.5-2keV}}/L_K = (3-8)\times10^{27} \ \rm{erg \ s^{-1} \ L_{K,\odot}}^{-1}$  \citep{sazonov06,revnivtsev08,bogdan11a}. By computing the $L_X/L_K$ ratios for the sample galaxies, we can determine whether the unresolved X-ray emission is dominated by faint compact and stellar sources or if other X-ray emitting components also play a role. 

To compute the $L_X/L_K$ ratios for each galaxy, we extracted X-ray energy spectra of the $D_{\rm{25}}$ regions with detected point sources excluded. The spectra were described with a two component model consisting of an optically-thin thermal plasma emission model (\textsc{APEC} in \textsc{Xspec}) and a power law model. The abundances were left free to vary, whereas the column density was fixed at the Galactic value \citep{dickey90}. From the  best-fit spectra, we computed the total $0.5-2$ keV  X-ray luminosity, which was used to derive the $L_X/L_K$ ratio for the  $D_{\rm{25}}$ regions.

However, the measured X-ray luminosities do not only reflect the luminosity of the undetected faint compact objects and possible hot gas, but are also contaminated by two other factors. First, a relatively small fraction ($\lesssim2\%$) of the detected source counts falls outside the source cells and contributes to the unresolved emission, which must be taken into account. Second, the available \textit{Chandra} exposures do not allow us to resolve all LMXBs. Therefore, the contribution of unresolved LMXBs must also be subtracted. To account for these contaminating factors, we follow the techniques described in \citet{bogdan11a}. The residual counts from the detected sources are subtracted on a source-by-source basis, whereas the contribution of unresolved LMXBs is removed using the average LMXB luminosity function \citep{gilfanov04}. 

The observed and contamination subtracted  $0.5-2$ keV band X-ray luminosities, the K-band luminosity, and the resulting $L_X/L_K$ ratios are listed in Table \ref{tab:list3}. In the same table we also list the ratios for the gas-free compact elliptical galaxy, M32 \citep{bogdan11a}, and the Solar neighborhood \citep{sazonov06}. The contamination-subtracted $L_X/L_K$ ratios obtained for NGC821 and NGC3379 are in the range of $(3-4)\times10^{27} \ \rm{erg \ s^{-1} \ L_{K,\odot}}^{-1}$,  in good agreement with the ratios observed for M32 and for the Solar neighborhood. However, in NGC4278 and NGC4697, we observe $(1.5-3.2)\times10^{28} \ \rm{erg \ s^{-1} \ L_{K,\odot}}^{-1}$,  a factor $\sim$$4-10$ times larger than that obtained for gas-free systems. Thus, based on the $L_X/L_K$ ratios, the unresolved emission in NGC821 and NGC3379 is dominated by the population of faint compact objects, whereas in NGC4278 and NGC4697, an additional X-ray emitting component is present, presumably originating from  X-ray gas. Based on the observed $0.5-2$ keV band $L_X/L_K$ ratios, we estimate that the contribution of faint unresolved sources is $\sim$$25\%$ in NGC4278 and $\sim$$10\%$ in NGC4697, implying that the hot X-ray gas is predominant at energies below $\sim$$2$ keV.

\begin{figure}[t]
    \leavevmode\epsfxsize=8.5cm\epsfbox{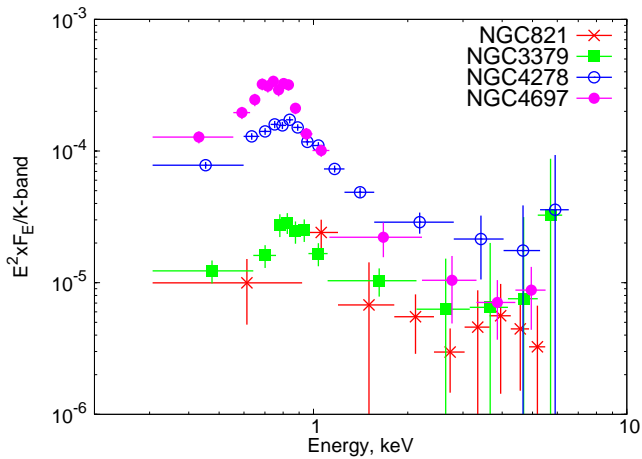}
    \caption{X-ray energy spectra of the four studied galaxies obtained for the $D_{\rm{25}}$ regions. To facilitate comparison, the spectra are normalized to the K-band luminosity of $10^{11} \ \rm{L_{K,\odot}}$ and to the distance of $10$ Mpc. Bright point sources are excluded and the background components are subtracted. }
\label{fig:spectra}
\end{figure}

\begin{figure*}[t]
  \begin{center}
    \leavevmode
      \epsfxsize=6in\epsfbox{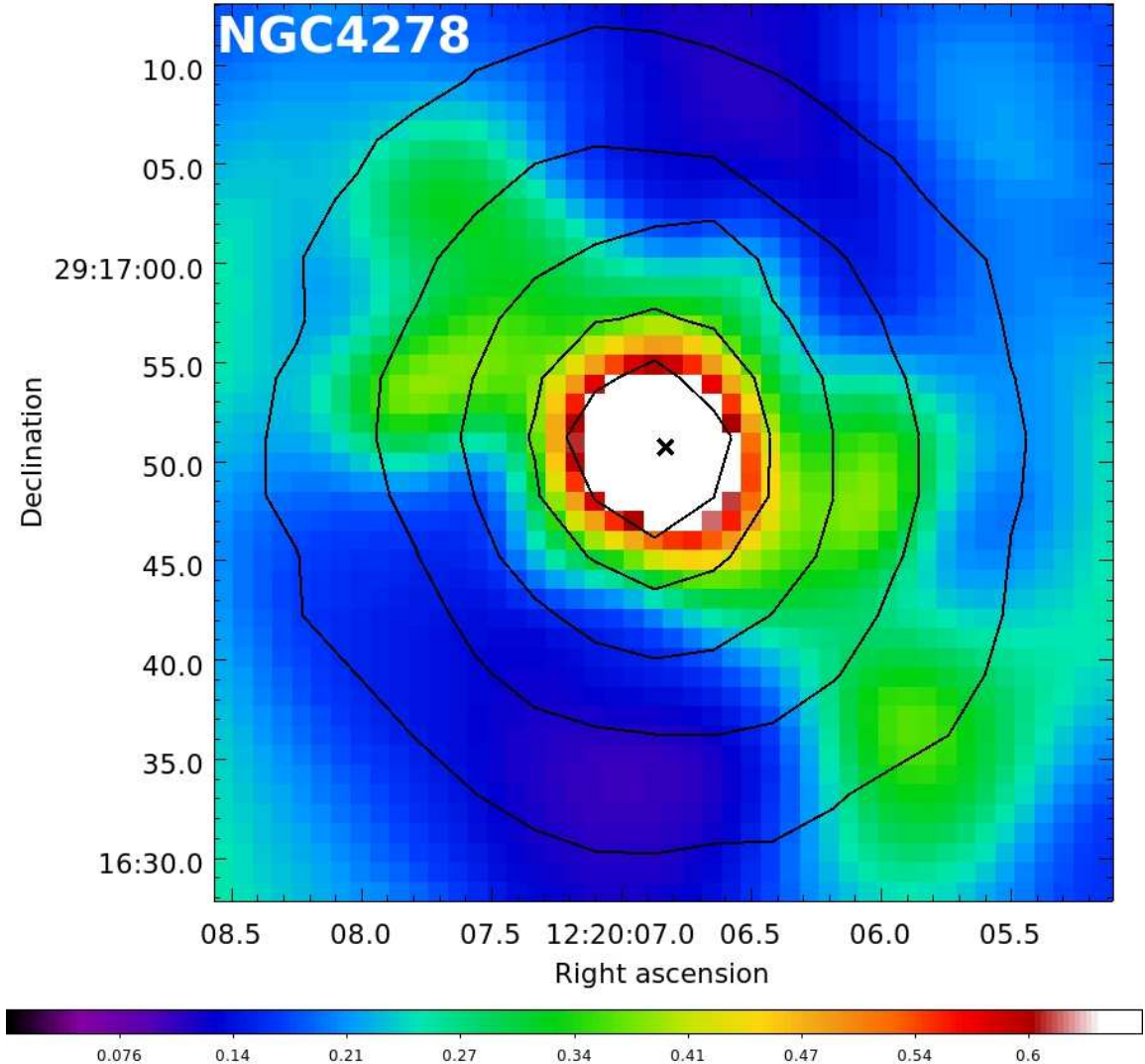}
      \caption{X-ray to K-band ratio image for the central $45\arcsec \times 45\arcsec$ region of NGC4278. Overplotted contours show the distribution of K-band light and the cross marks the center of the galaxy. The hot gas exhibits a bipolar morphology in the  norteast-southwest direction out to at least $\sim$$20\arcsec$ ($\sim$$1.6$ kpc) projected distance, thereby  hinting at the presence of a galactic-scale outflow. Details on how the image was constructed are discussed in Section \ref{sec:morphology}. }
\vspace{0.5cm}
     \label{fig:ratio_ngc4278}
  \end{center}
\end{figure*}

\begin{figure*}[t]
  \begin{center}
    \leavevmode
      \epsfxsize=6in\epsfbox{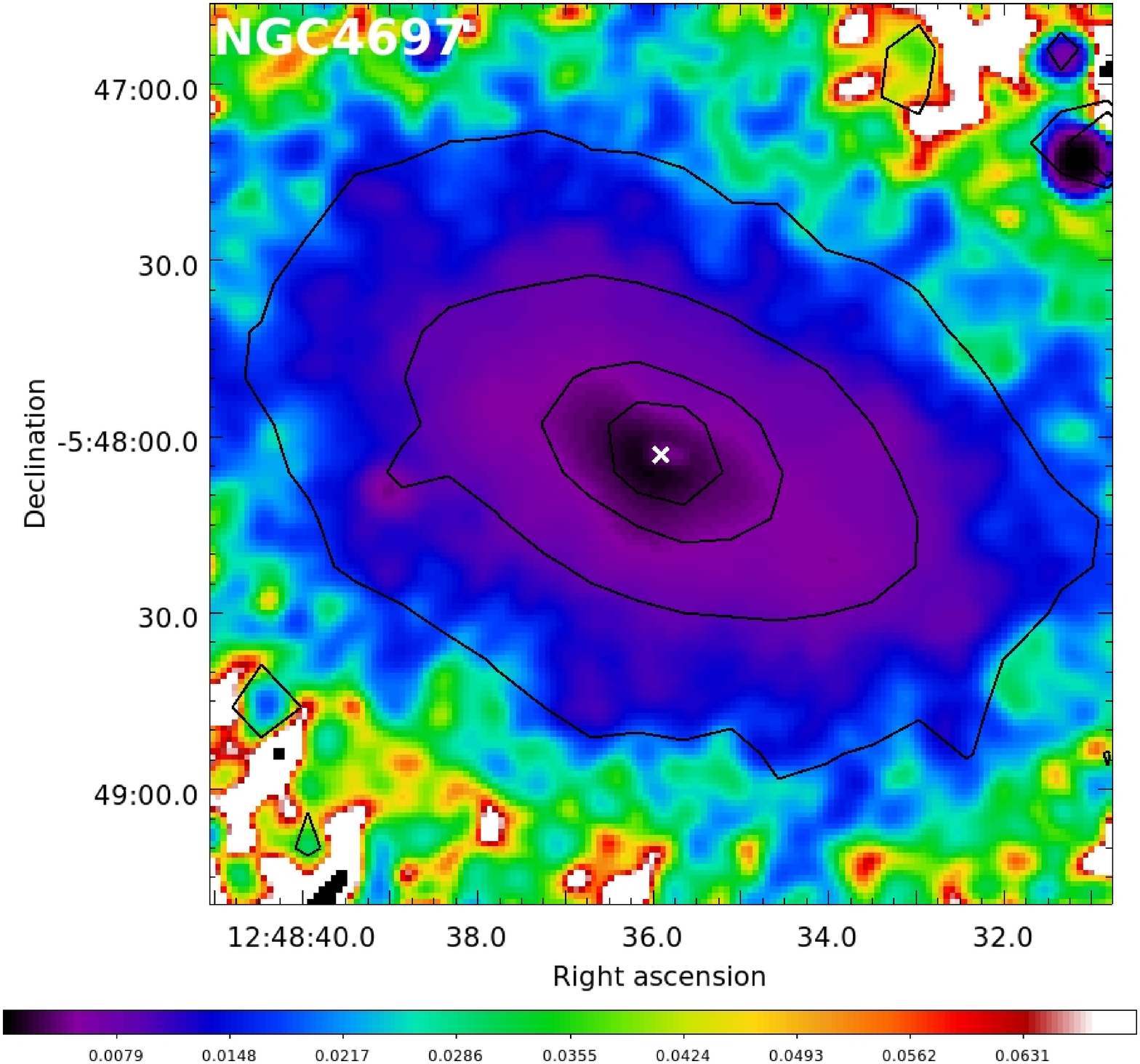}
      \caption{X-ray to K-band ratio image for the central $150\arcsec \times 150\arcsec$ region of NGC4697.  Overplotted contours show the distribution of K-band light and the cross marks the center of the galaxies. The image illustrates that the hot X-ray gas has a notably broader distribution than the stellar light. However, besides its broad distribution, the X-ray gas  does not exhibit any further asymmetries. Details on how the image was constructed are discussed in Section \ref{sec:morphology}. }
\vspace{0.5cm}
     \label{fig:ratio_ngc4697}
  \end{center}
\end{figure*}

\subsection{X-ray energy spectra}
\label{sec:spectra}
To confirm the presence of hot X-ray gas in NGC4278 and NGC4697, we extracted their X-ray energy spectra and compared them to those derived for the gas-free galaxies NGC821 and NGC3379. The spectra, shown in Figure \ref{fig:spectra}, are extracted from within the $D_{\rm{25}}$ regions. To facilitate the comparison of the observed spectra, we normalize the spectra to the K-band luminosity of $10^{11} \ \rm{L_{K,\odot}}$ and  the distance of $10$ Mpc. Bright point sources are excluded and all background components are subtracted from the depicted spectra. 

The spectra demonstrate that NGC4278 and NGC4697 host a prominent soft component at energies $\lesssim1.5$ keV, whereas this component is about an order-of-magnitude weaker in NGC821 and NGC3379. We stress that the weak soft component in the spectra of NGC821 and NGC3379 is most likely due to the soft X-ray spectra of ABs \citep[e.g.][]{revnivtsev08}.  Above $\sim$$1.5$ keV energy, the spectra are in fairly good agreement with each other, which is also confirmed by the $2-10$ keV band $L_X/L_K$ ratios \citep{bogdan11a}. 

To characterize the physical properties of the hot gas, we use two approaches. As a first approach, we fit the spectra with a simple two component model, consisting of an optically-thin thermal plasma emission model (\textsc{APEC} in \textsc{Xspec}) and a power law component. During the fitting procedure, the column density is fixed at the Galactic value (Table \ref{tab:list2}), whereas the abundance is set free. In NGC4278, the best-fit temperature and abundances of the \textsc{APEC} models are $kT=0.46\pm0.02$ keV and $0.12^{+0.11}_{-0.03}$ of the Solar values \citep{grevesse98}. Using the same model for NGC4697, the best-fit temperature and abundances of the thermal model are $kT=0.34\pm0.03$ keV and $0.14^{+0.06}_{-0.04}$ of  Solar  \citep{grevesse98}.  This simple two component model gives an acceptable fit for the X-ray spectra of both galaxies, thereby confirming that the notable fraction of the soft band emission originates from hot gas. However, this two component model does not take into account that a small but certain fraction of the soft band emission arises from the population of unresolved faint compact objects. 

As a second approach, we use the spectrum of NGC3379 as a template to describe the X-ray emission originating from faint unresolved sources. The X-ray spectrum of NGC3379 can be described with the combination of a thermal and a power law component, where the \textsc{APEC} model has a temperature of $0.67$ keV and Solar abundance, whereas the slope of the power law is $\Gamma=1.7$. We employ this template spectrum and rescale it by the K-band luminosity within the spectral extraction regions to account for the X-ray emission associated with ABs and CVs in NGC4278 and NGC4697. Thus, the spectrum of NGC4278 and NGC4697 is fit with a four component model, which comprises two thermal and two power law models. One thermal and one power law component describes the emission arising from ABs and CVs, therefore the parameters of these models are fixed at the values deduced from the NGC3379 spectrum and are scaled according to the K-band light of the given regions. The second power law component  accounts for the emission originating from unresolved LMXBs in NGC4278 and NGC4697. We emphasize that in the soft band the emission associated with the power law models does not play a dominant role. Indeed, in the $0.3-1.2$ keV band,  $\lesssim25\%$ of the total flux originates from the power law components. The parameters of the hot gas associated with NGC4278 and NGC4697 are described with the second thermal model. Interestingly, the derived best-fit gas temperature of NGC4278 is $0.34\pm0.03$ keV, which is significantly lower than that obtained with the simple  two component models and is in good agreement with that measured by \citet{pellegrini12}. In the case of NGC4697, the best-fit temperature is $0.32\pm0.02$ keV, being in good agreement with the previously measured value. We also note that the best-fit abundances remained essentially unchanged, that is, both NGC4278 and NGC4697 exhibit strongly sub-solar abundances. 

The difference in the observed gas temperatures in NGC4278 is most likely due to its gas-poor nature. Indeed, in NGC4278, $\sim$$25\%$ of the $0.5-2$ keV band luminosity arises from the population of CVs and ABs, which adds a notable contribution to the observed thermal emission. Therefore, when fitting the X-ray spectrum with a simple two component model consisting of one thermal plasma model and one power law model, the best-fit temperature of the thermal model is the weighted average temperature of the truly diffuse gaseous emission and the unresolved faint compact objects (for further discussion see Section \ref{sec:metal}). When two distinct thermal models are applied, their temperatures are not averaged, which in the case of NGC4278 results in a lower temperature for the truly diffuse gaseous emission. Obviously, this effect is less significant in more gas-rich galaxies, such as NGC4697, where the emission from  unresolved CVs and ABs plays a lesser role. In further discussions, we employ the above-discussed four component spectral model, that is we take into account the contribution of faint unresolved sources to the X-ray emission.

\begin{figure*}[t]
  \begin{center}
    \leavevmode
      \epsfxsize=8.5cm\epsfbox{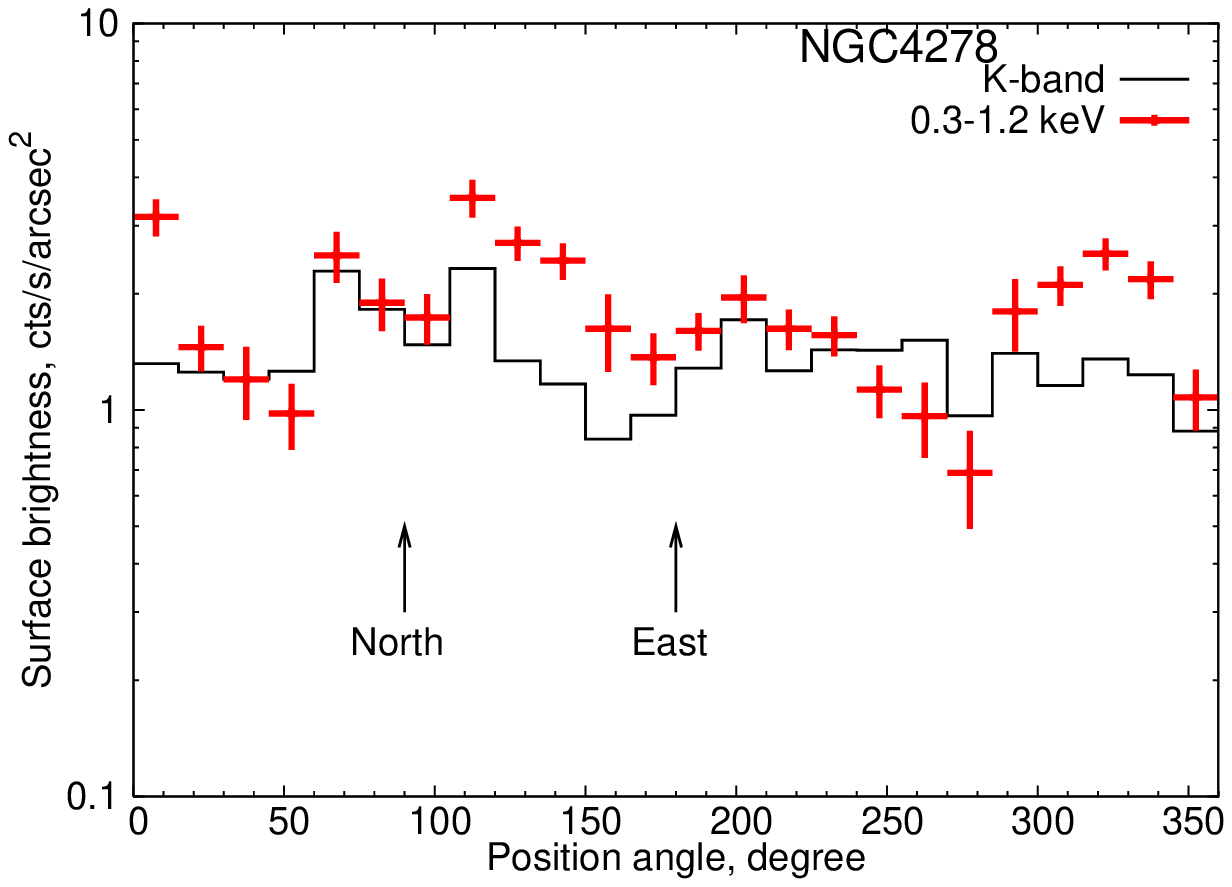}
\hspace{0.4cm} 
      \epsfxsize=8.5cm\epsfbox{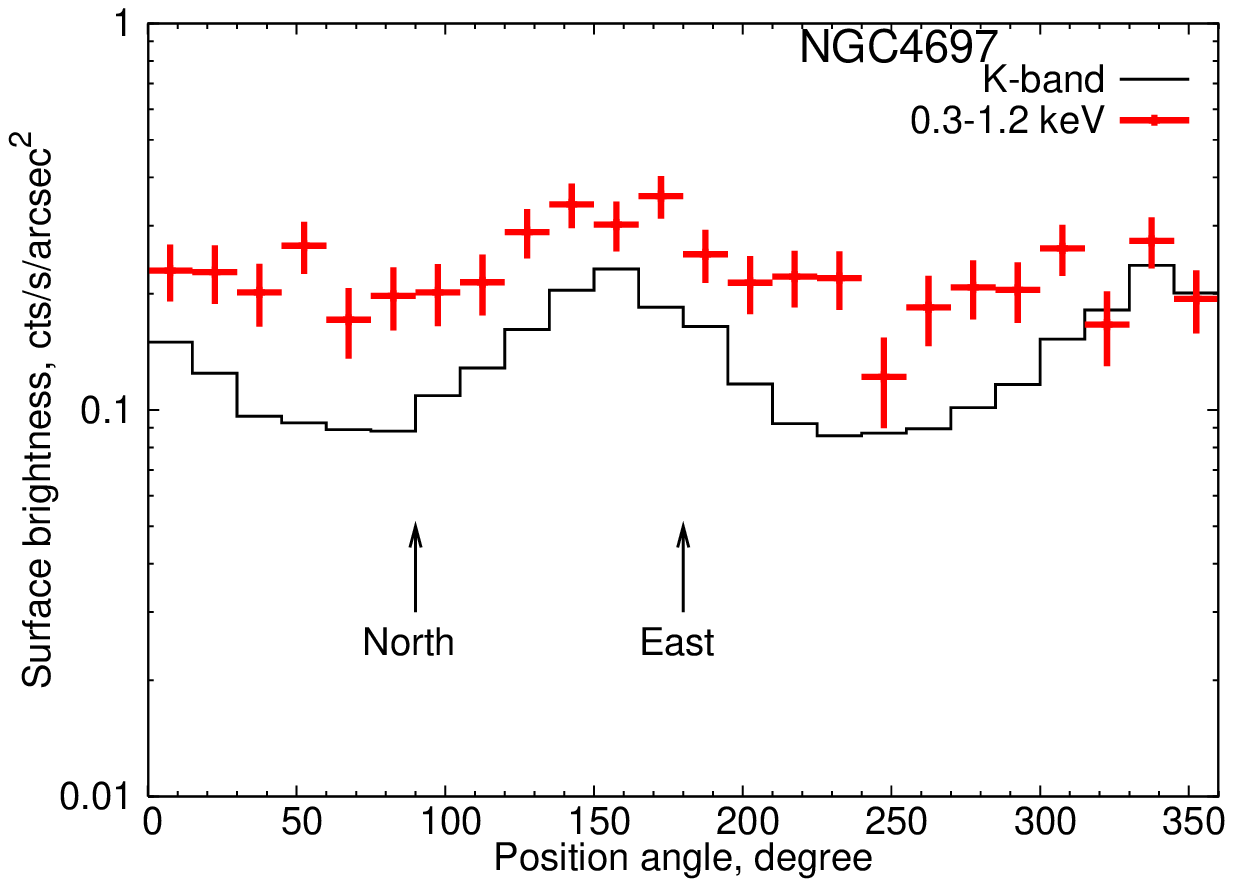}
      \caption{Surface brightness distribution of the $0.3-1.2$ keV energy band unresolved X-ray emission for NGC4278 (left panel) and NGC4697 (right panel). The profiles are obtained for circular wedges, whose inner and outer radii are $7-23 \arcsec$ for NGC4278 and $20-50 \arcsec$ for NGC4697. As shown in the plots, $90\degr$ corresponds to north and $180\degr$ corresponds to east. The X-ray peaks observed for NGC4278 in Figure \ref{fig:ratio_ngc4278} are located at $\sim$$135\degr$ (north-east) and $\sim$$315\degr$ (south-west).  Vignetting correction is applied and the background components are subtracted. The X-ray profiles are compared with the K-band profiles. Note that the normalization of the K-band profiles agree with those of Figure \ref{fig:profile}.}
\vspace{0.5cm}
     \label{fig:wedges}
  \end{center}
\end{figure*}

\subsection{The X-ray gas content of NGC3379}
In NGC3379, we did not detect notable X-ray emission from hot ionized gas, in agreement with the results of \citet{revnivtsev08}. However, our conclusions are in conflict with those of \citet{trinchieri08}, who claimed that NGC3379 hosts X-ray emitting gas within the central $<20\arcsec$. In principle, it is possible that the applied point source exclusion regions by \citet{trinchieri08} were too small, and the excess emission in the central regions is due to ``spill-over'' counts from bright point sources. Within the central $20\arcsec$ circular region, we detect $12809$ source counts in the $0.5-2$ keV band, of which $12491$ are associated with bright point sources, if $2\arcsec$ aperture radii are used as in \citet{trinchieri08}. That is, only $318$ counts, or $\approx$$2.5\%$ of the total counts, could be associated with the diffuse emission within the central $20\arcsec$ region. Since, on-axis, the applied $2\arcsec$ source aperture encircles $\sim$$98\%$ of the point spread function for $0.5$ keV energy, the number of  ``spill-over'' counts from point sources is $\sim$$250$ within the central $20\arcsec$. Thus, $\sim$$79\%$ of the unresolved  counts are likely to be associated with bright point sources rather than truly diffuse emission. Finally,  we also mention that $\approx$$43\%$ of the source counts originate from the nuclear source in NGC3379, hence the fraction of  ``spill-over'' counts is most significant in the central $\sim$$10\arcsec$ region. Therefore, to avoid being dominated by the source counts falling outside the source cells, we excluded the central $7\arcsec$ of NGC3379 throughout this study.

\begin{figure*}[t]
  \begin{center}
    \leavevmode
      \epsfxsize=8.5cm\epsfbox{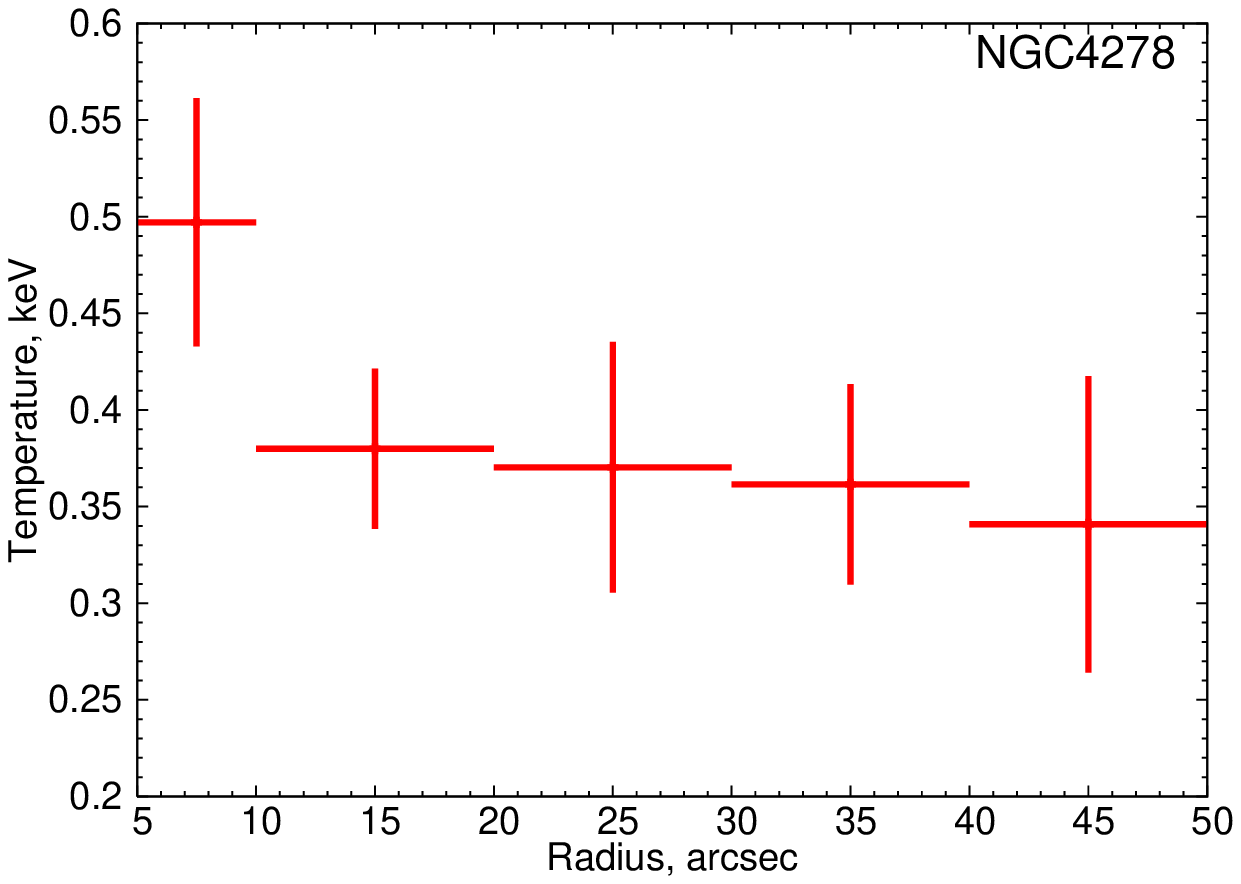}
\hspace{0.4cm} 
      \epsfxsize=8.5cm\epsfbox{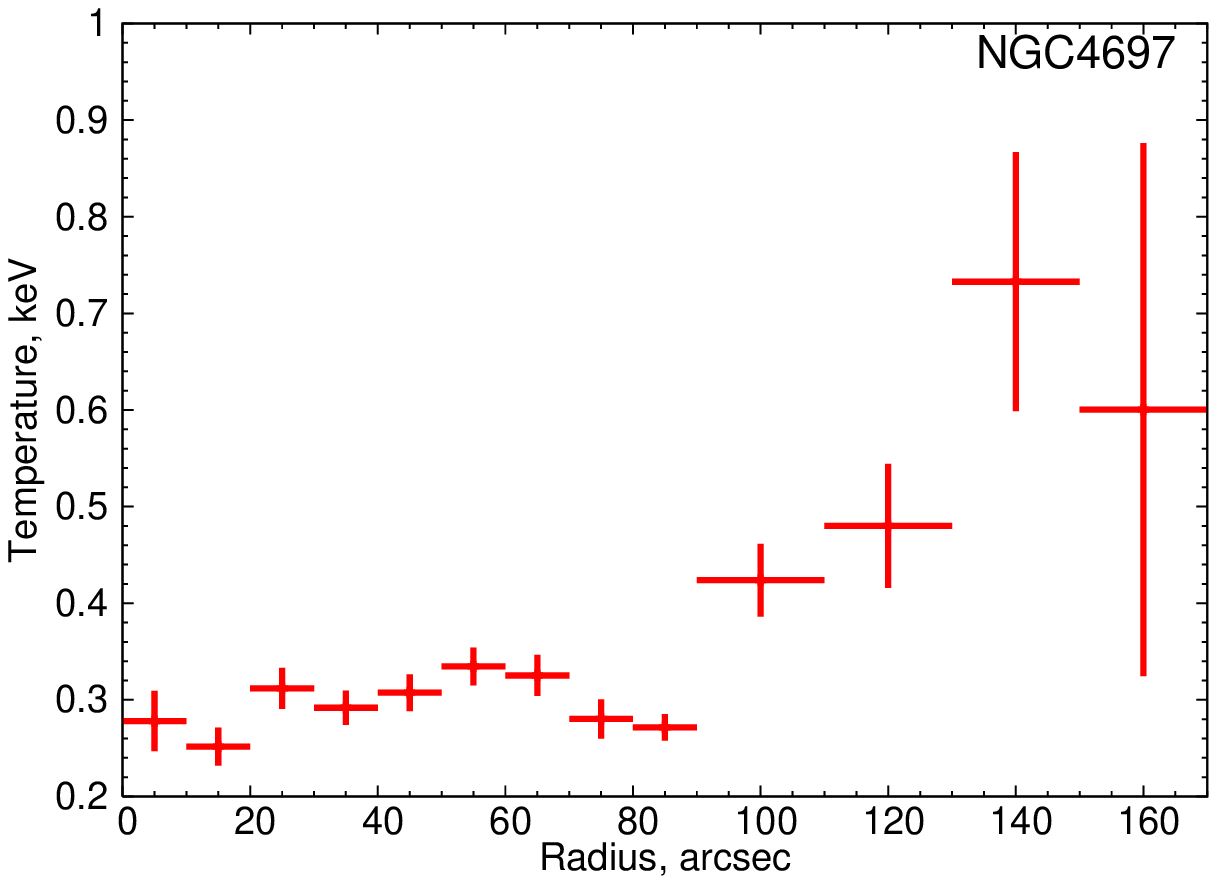}
      \caption{Projected temperature profiles of the diffuse emission in NGC4278 (left panel) and NGC4697 (right panel). To obtain the profiles, the X-ray energy spectra were extracted from circular annuli centered on the center of the galaxies. The spectra were described with a four component model consisting of two thermal plasma (\textsc{APEC} in Xspec) and two power law models. The parameters of one thermal and one power law component are determined from the template spectrum of NGC3379, which was used to describe the emission associated with the populations of faint compact objects.  The depicted temperatures are the best-fit temperatures of the second thermal component. The abundance was set to $0.1$, which is close to that obtained for the spectrum integrated over the entire galaxy.}
\vspace{0.5cm}
     \label{fig:temperature}
  \end{center}
\end{figure*}

\subsection{Morphology of the hot X-ray gas \\ in NGC4278 and NGC4697}
\label{sec:morphology}
Having demonstrated the presence of hot gas in NGC4278 and NGC4697, we intend to map its morphology. Understanding the spatial distribution of the gas may provide important hints about its physical state. We use two techniques to determine the morphology of the gas. First, we produce a ratio image of the $0.3-1.2$ keV band X-ray image and the K-band image. The X-ray-to-K-band ratio image allows us to only study the distribution of the hot gas, since other X-ray emitting components -- whose luminosity is proportional with the K-band  light -- are removed. However, this procedure has certain weaknesses: in regions where the X-ray emission is not associated with the K-band light, the ratio image may have very large values; and due to the strongly processed nature of the images, the central few arcsec regions cannot be trusted. To gain insight into the distribution of the hot gas, we prepared both the X-ray and the K-band images. In particular, we excluded the resolved X-ray sources from the  \textit{Chandra} images, and filled their locations with the local emission level applying the \textsc{dmfilth} task of \textsc{CIAO}. Moreover, the X-ray images are vignetting corrected, the estimated background level is subtracted, and adaptive smoothing is applied using the \textsc{CIAO} \textsc{csmooth} tool. The K-band images were smoothed by convolving them with a Gaussian. The applied smoothing widths are comparable to the typical smoothing widths near the center of the X-ray images. The obtained ratio images are renormalized, and the highest pixel value is assigned the value of $1$.

The X-ray-to-K-band  ratio images for NGC4278 and NGC4697 are depicted in Figure \ref{fig:ratio_ngc4278} and Figure \ref{fig:ratio_ngc4697}, respectively. For NGC4278 we show the central $45\arcsec \times 45\arcsec$ region, where the hot X-ray emitting gas is most prominent.  Figure \ref{fig:ratio_ngc4278} demonstrates that the gas distribution in NGC4278 significantly deviates from the stellar light: the X-ray emission is strongly elongated in the northeast-southwest direction, which may indicate the presence of a galactic-scale outflow.  Figure \ref{fig:ratio_ngc4697} shows the X-ray-to-K-band ratio image for NGC4697. Since the X-ray gas has a broad distribution in NGC4697, we studied the central $150\arcsec\times150\arcsec$ region. We note that $\sim$$80\%$ of the stellar light is confined within this region. The ratio image confirms that the hot gas is significantly more extended than the stellar light. However, besides the broad gas distribution, we did not detect any notable asymmetries associated with the gas distribution. 

As a second approach, we produce X-ray surface brightness distribution profiles of the $0.3-1.2$ keV band unresolved emission using circular wedges (Figure \ref{fig:wedges}). The profiles are obtained in the same way as described in Section \ref{sec:profiles}. The X-ray profiles are compared with the K-band light distributions, whose normalization agrees with those applied in Figure \ref{fig:profile}. The major advantage of this method is that we deal with the original X-ray and K-band images, which are not affected by the various processing procedures. Therefore, these profiles can confirm our findings based on the X-ray-to-K-band ratio images. Since the hot gas is centrally concentrated in NGC4278, the inner and outer radii of the applied annulus is   $7-23\arcsec$. The left panel of Figure \ref{fig:wedges} confirms the bipolar distribution of the hot ionized gas: the soft band X-ray emission has two major peaks, namely at the position angles of $105-165\degr$ and $285-345\degr$, which correspond to the northeast and southwest directions. In the right panel of Figure \ref{fig:wedges}, we show the surface brightness profiles for NGC4697, which were obtained for a circular annulus with inner and outer radii of   $20-50\arcsec$. As expected from the X-ray-to-K-band image, the X-ray profiles are approximately uniformly higher than the K-band profiles at every position angle. Thus, the profiles do not unveil any major asymmetries associated with the X-ray light distribution.

\section{Discussion}
\subsection{Physical state of the hot gas in NGC4278}
\label{sec:outflow_ngc4278}
The bipolar distribution (Figure \ref{fig:ratio_ngc4278}) of the hot gas in NGC4278 suggests that it is not in hydrostatic equilibrium, but more likely in an outflow state. In principle, galactic-scale outflows can be driven by the energy input of SNe Ia and the outflowing mass can be replenished by stellar yields from evolved stars \citep{david06}. To explore whether a galactic-scale outflow can be sustained in NGC4278, we investigate the mass and energy budget of the galaxy. 

Since the X-ray gas is most luminous in the central regions of the galaxy, we use two rectangular boxes with $9\arcsec\times20\arcsec$ ($ 0.7 \times 1.56 $ kpc) that are coincident with the elongated features detected in the X-ray-to-K-band image (Figure \ref{fig:ratio_ngc4278}). For further computations, we assume that the path length of the gas is $0.7$ kpc. We extract a combined spectrum from these regions, which also confirms the thermal nature of the emission. The best-fit temperature and abundance of the hot gas is $kT=0.43\pm0.03$ keV and $0.12^{+0.11}_{-0.03}$ Solar \citep{grevesse98}. The $0.5-2$ keV X-ray luminosity of the thermal emission in the studied cylindrical shell is $3\times10^{38} \ \rm{erg \ s^{-1}}$.  From the emission measure, we estimate that the gas mass within the studied volume is $\sim$$10^6 \ \rm{M_{\odot}}$, the average number density is $0.05 \ \rm{cm^{-3}}$, and the cooling time is $t_{\rm{cool}}= (3kT)/(n_e  \Lambda(T))\approx 1.2\times10^8$ years. 

Given the parameters of the X-ray gas, we investigate whether the stellar yields from evolved stars can maintain the outflow. The average stellar yield in elliptical galaxies is $ \sim$$ 0.0021 \ (L_{K}/ L_{K,\sun}) \ \mathrm{M_{\sun} \ Gyr^{-1}} $ \citep{knapp92}. The K-band luminosity of the cylindrical shell is $ L_{K} = 6.0 \times 10^{9} \ \mathrm{L_{K,\sun}}$, and hence the mass loss rate is $\dot{M}_{\star} = 0.013 \ \rm{M_{\odot} \ yr^{-1}}$. Thus, the stellar yields are able to produce the observed gas mass in a timescale of $\sim$$8\times10^7$ years, which is shorter, but comparable with the cooling time of the gas. The observed low gas mass and the short replenishment time of the gas indicates that the stellar yields from evolved stars are not retained in the galaxy, but are removed in the form of a galactic-scale outflow.

\begin{figure*}[t]
  \begin{center}
    \leavevmode
      \epsfxsize=8.5cm\epsfbox{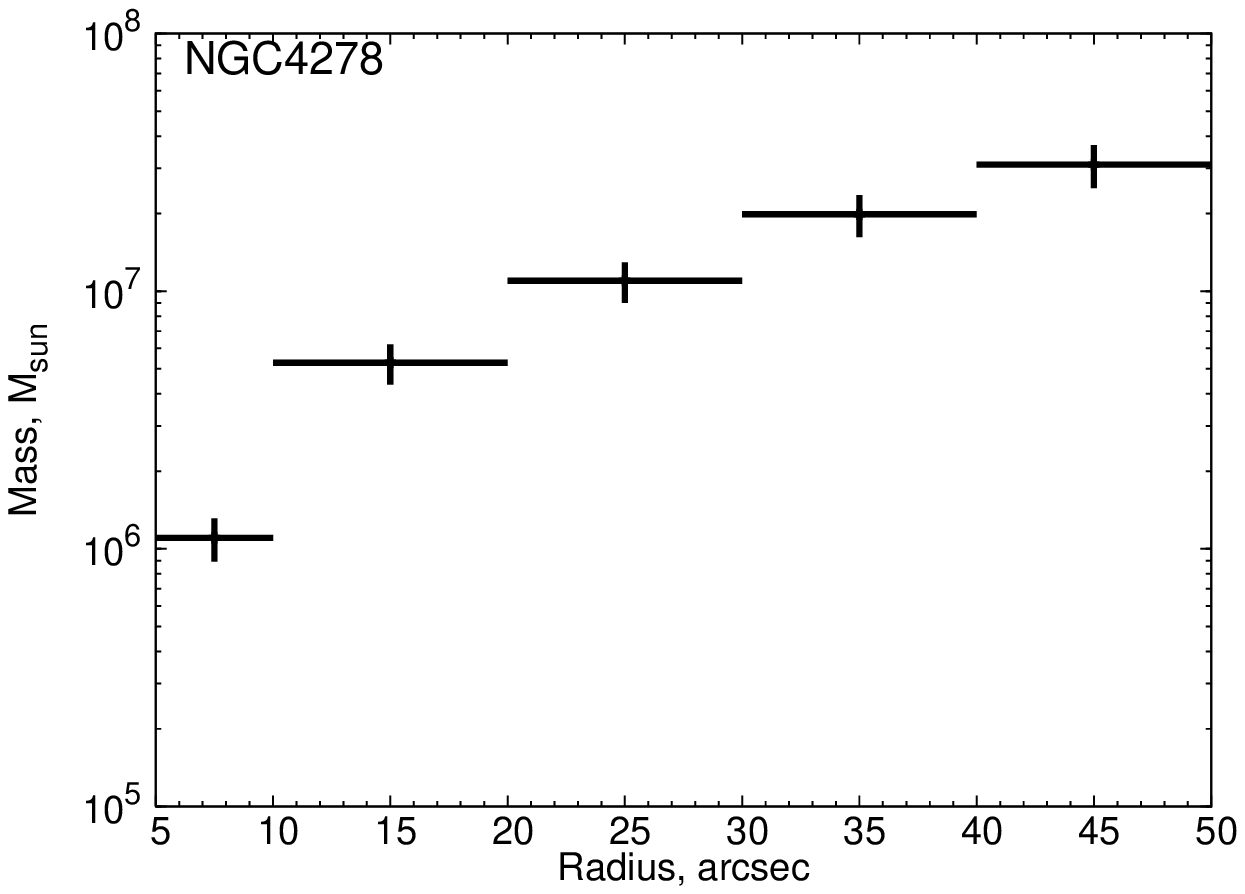}
\hspace{0.4cm} 
      \epsfxsize=8.5cm\epsfbox{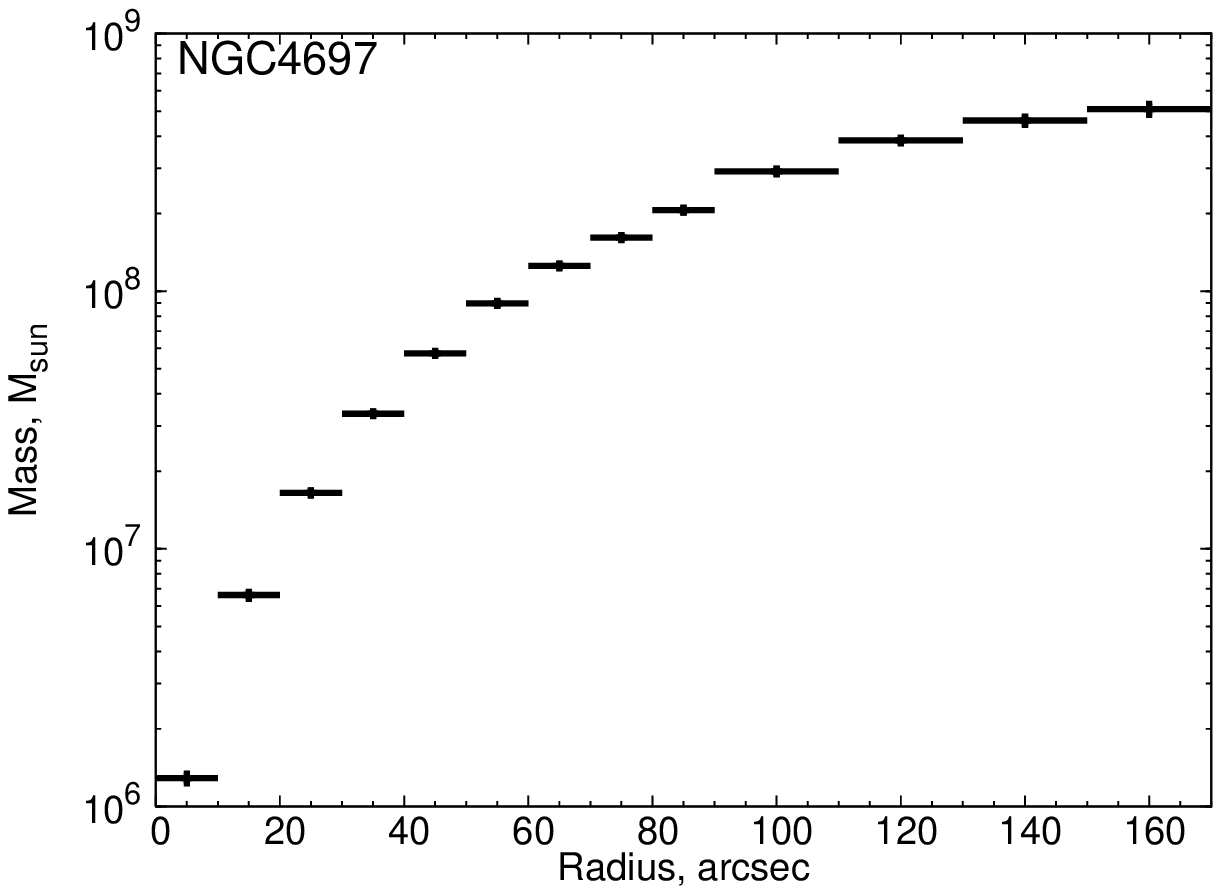}
      \caption{Cumulative gas mass distribution for NGC4278 (left panel) and NGC4697 (right panel). To obtain the gas mass profiles, spherical symmetric gas distribution was assumed. The total stellar mass of NGC4278 within $50\arcsec$ is $4.9\times10^{10} \ \rm{M_{\odot}}$, hence the X-ray gas mass fraction is $\sim$$0.06\%$. NGC4697 is a significantly more gas-rich system -- within $90\arcsec$  its total stellar mass is $5.4\times10^{10} \ \rm{M_{\odot}}$, implying a gas mass fraction of $\sim$$0.4\%$.}
\vspace{0.5cm}
     \label{fig:massprofile}
  \end{center}
\end{figure*}

If the gas is outflowing from the galaxy, SNe Ia are likely candidates to drive the outflow. Therefore, we compute whether the energy input from SNe Ia is sufficient to sustain the outflow. The frequency of SNe Ia in old stellar populations, such as NGC4278, is proportional to the stellar mass and hence to the K-band luminosity of the galaxy. Indeed, \citet{mannucci05} observed an SN Ia frequency of $N_{\mathrm{SNIa}} = 0.035^{+0.013}_{-0.011} \ \mathrm{SNu} $ for E and S0 galaxies, where $ 1 \ \mathrm{SNu} = 1 \ \mathrm{SN}/10^{10}\  L_{K,\sun} $ per century. Based on the K-band luminosity of the selected regions, the estimated average  SN Ia rate is $2.1 \times 10^{-4} \ \rm{yr^{-1}}$. Assuming that the average energy output of one SN Ia is $10^{51} \ \rm{erg}$, we estimate that at most $ 6.7 \times 10^{39} \ \mathrm{erg \ s^{-1}} $ is available to heat the gas and drive the outflow. The required power to drive the outflow can be computed as 
\begin{eqnarray}
P_{\rm{lift}}=7.2\dot{M}_{\star} \sigma^2
\end{eqnarray}
where $\dot{M}_{\star}$ is the stellar mass loss rate and $\sigma$ is the velocity dispersion of the galaxy \citep{david06}. Using the velocity dispersion of NGC4278 ($\sigma=237.2 \ \rm{km \ s^{-1}}$ -- \citealp{paturel03}; HyperLeda\footnote{http://leda.univ-lyon1.fr}) and the stellar mass-loss rate ($\dot{M}_{\star} = 0.013 \ \rm{M_{\odot} \ yr^{-1}}$), we estimate that $P_{\rm{lift}}=3.3\times 10^{39} \ \rm{erg \ s^{-1}}$ is needed to drive the outflow. The gas shed by evolved stars is shock heated to the virial temperature of $kT=\sigma^2 \mu m_p = 0.36 $ keV, adopting $\mu=0.62$. This value is approximately consistent with the observed X-ray gas temperature. Therefore, the total power required to drive the outflow is $P_{\rm{tot}} \sim 3.3 \times 10^{39} \ \mathrm{erg \ s^{-1}} $, which is $\sim$$50\%$ of that available from SNe Ia. Thus, we confirm that SNe Ia are capable of sustaining a galactic-scale outflow in NGC4278. 

Although the energy input of SNe Ia can sustain a galactic-scale wind, in principle it is possible that the central supermassive black hole also plays a role in driving the outflow. NGC4278 hosts a low-luminosity active galactic nucleus (AGN), which at present is in outburst, exhibiting a parsec scale radio jet \citep{giroletti05}. The estimated black hole mass of NGC4278 based on  the $M_{\bullet}-\sigma $ relation \citep{gultekin09} and the velocity dispersion of the galaxy ($\sigma=237.2 \ \rm{km \ s^{-1}}$) is $\sim$$3.3 \ \times{10^8} \ \rm{M_{\odot}}$. Assuming that the outflow is driven by the energy input of a preceding AGN outburst, we estimate the occurrence time of the presumed outburst. Based on the X-ray surface brightness profile (Figure \ref{fig:profile}), the outflow may extend to $\sim$$40\arcsec$ ($\sim$$3.1$ kpc). If the gas moves with the sound speed, which in a $0.4$ keV plasma is $c_s=\sqrt{(\gamma kT)/(\mu m_H})\approx320 \ \rm{km \ s^{-1}}$, then adopting $\gamma=5/3$ and $\mu=0.62$, we derive that the outburst occurred $\sim$$9.5\times10^6 $ years ago. However,  two observational flaws make this proposed interpretation unlikely. First, if an outburst happened  $\sim$$9.5\times10^6 $ years ago, the detection of radio emission associated with the outburst would be likely. However, in the  1.4 GHz Very Large Array observation of NGC4278, only the nucleus is detected, but no extended radio emission is observed. Second, AGN outbursts produce X-ray cavities in the hot gas distribution  \citep[e.g.,][]{bogdan11b}. Our \textit{Chandra} images do not reveal such cavities. Instead, the morphology of the X-ray gas is consistent with a bipolar galactic-scale outflow (Figure \ref{fig:ratio_ngc4278}).   We thus conclude that the galactic-scale outflow in NGC4278 is most likely driven by the energy input of the population of SNe Ia. 

Based on the detection of an SN Ia driven outflow in NGC4278, we can constrain the supernova efficiency parameter ($\xi$), which is the percentage of the supernova energy that heats the gas and drives the outflow. Numerical simulations by \citet{thornton98} showed that the value of $\xi$ is typically $\sim$$10\%$, but may be as high as $\sim$$30\%$. The presence of an SN Ia driven outflow in NGC4278 implies that the supernova efficiency parameter is  $\xi\gtrsim50\%$, exceeding those predicted by numerical simulations. 

To describe the large-scale properties of the  X-ray gas, we extract spectra using circular annuli centered on the centroid of NGC4278. The column density is fixed at the Galactic value, while the abundance of the second thermal model is fixed at $0.1$ Solar, which is close to the best-fit average value. The projected temperature profile is shown in the left panel of Figure \ref{fig:temperature}, which illustrates that the hot gas is approximately isothermal with $kT\sim0.35$ keV. To estimate the gas mass within NGC4278, we assume spherical symmetric gas distribution. In the left panel of Figure \ref{fig:massprofile} we depict the cumulative gas distribution of NGC4278, which reveals that within $50\arcsec$ ($\sim$$3.9$ kpc) it confines $\sim$$3\times10^7 \ \rm{M_{\odot}}$ gas. Within this region, the total stellar mass is $\sim$$4.9\times10^{10} \ \rm{M_{\odot}}$, implying a gas mass fraction of $\sim$$0.06\%$.

\subsection{Physical state of the hot gas in NGC4697}
\label{sec:gas_ngc4697}
The extended spatial distribution of the hot ionized gas in NGC4697 is consistent with two physical states. The X-ray gas may either  be in hydrostatic equilibrium in the potential well of the galaxy or it may undergo a subsonic SN Ia driven outflow. To discriminate between these possibilities, we compute the physical properties of the X-ray gas and its mass and energy budget. 

To explore the physical properties of the hot X-ray gas in NGC4697, we extract X-ray energy spectra using  circular annuli centered on the centroid of the galaxy.  Since the average abundance within the $D_{\rm{25}}$ ellipse is $0.14^{+0.06}_{-0.04}$ \citep{grevesse98}, we fix the abundances of the second thermal model at $0.1$ Solar. The column density is fixed at the Galactic value (Table \ref{tab:list2}). From the best-fit temperatures of the thermal components, we derive a projected temperature profile, shown in the right panel of Figure \ref{fig:temperature}. The profile demonstrates that within the central $\sim$$90\arcsec$, the X-ray gas is approximately isothermal with $kT\sim0.3$ keV, beyond which radius the X-ray gas temperature gradually increases to $kT\sim0.7$ keV. This result can be interpreted in a simple manner in view of the environment of NGC4697. Indeed, NGC4697 is a member of a small galaxy group with $10$ members, hence the gas with $\sim$$0.7$ keV temperature is most likely associated with a larger scale group atmosphere. Thus, within the central $\sim$$90\arcsec$ radius, the $\sim$$0.3$ keV galaxy gas dominates the diffuse emission, and at larger radii the hotter group gas becomes predominant.  

We derive the X-ray gas mass by employing the same spectral extraction regions, and assuming a spherically symmetric gas distribution.  From the emission measure of the thermal component, we derive the confined gas mass in each shell. The resulting cumulative mass profile is depicted in the right panel of Figure \ref{fig:massprofile}, which shows that within $90\arcsec$ ($\sim$$5.1$ kpc),  NGC4697 contains $\sim$$2.1 \ \times{10^8} \ \rm{M_{\odot}}$ of gas. We note that this region comprises $\sim$$82\%$ of the total stellar light of the galaxy. The total stellar mass within $90\arcsec$ is $5.4\times10^{10} \ \rm{M_{\odot}}$, hence the X-ray gas mass fraction is $\sim$$0.4\%$, which significantly exceeds that of NGC4278 ($\sim$$0.06\%$; Section \ref{sec:outflow_ngc4278}) and is consistent with values obtained for more X-ray luminous galaxies ($\sim$$1\%$; \citealp{fukazawa06}), whose X-ray gas content is in hydrostatic equilibrium. This suggests that the X-ray gas content of NGC4697 is in hydrostatic equilibrium rather than in an outflow state.

To further study the physical state of the gas, we derive the timescale on which the observed gas can be replenished by the stellar yields of evolved stars. Since the galaxy gas dominates the X-ray emission within $\sim$$90\arcsec$ radius, the calculations presented below refer to this region. Using the average stellar yields for elliptical galaxies \citep{knapp92}, we estimate an average mass loss rate of  $0.14 \ \rm{M_{\odot} \ yr^{-1}}$. Hence, evolved stars can replenish the observed $\sim$$2.1 \ \times{10^8} \ \rm{M_{\odot}}$ X-ray gas mass on a timescale of $\sim$$1.5$ Gyr. This timescale exceeds the cooling time, which is $\lesssim1$ Gyr within the central $\sim$$90\arcsec$  region. Thus, most, if not all, hot gas has been retained in hydrostatic equilibrium in the past $\sim$$1.5$ Gyr, implying the absence of galactic-scale outflows.

Alternatively, it is also feasible that only a fraction of the X-ray gas leaves the galaxy in a subsonic outflow driven by the energy input of SNe Ia \citep{david91}. We derive the available energy input from SNe Ia, using the average SN Ia frequency in E/S0 galaxies \citep{mannucci05} and the K-band luminosity of NGC4697. We obtain an SN Ia rate of  $2.3\times10^{-3} \ \rm{yr^{-1}}$. Assuming that the average energy output from SNe Ia is $10^{51} \ \rm{erg}$, SNe Ia can add at most $ 7.3 \times 10^{40} \ \mathrm{erg \ s^{-1}} $. To estimate the required power to drive the outflow, we use the velocity dispersion of NGC4697 ($\sigma=170.9 \ \rm{km \ s^{-1}}$ -- \citealp{paturel03}; HyperLeda\footnote{http://leda.univ-lyon1.fr}), the stellar mass-loss rate ($\dot{M}_{\star} = 0.14 \ \rm{M_{\odot} \ yr^{-1}}$), and Equation (2). We thus obtain  $ P_{\rm{lift}} = 1.9 \times 10^{40} \ \mathrm{erg \ s^{-1}} $. If the gas is heated to the virial temperature of $kT=0.19$ keV, then SNe Ia need to heat the gas an additional $\sim$$0.1 $ keV per particle, which in total corresponds to $\sim$$ 10^{39} \ \mathrm{erg \ s^{-1}} $.  Hence, the total required power to heat the gas and drive an outflow is  $\sim$$ 2 \times 10^{40} \ \mathrm{erg \ s^{-1}} $, which is $\sim$$30\%$ of that available from SNe Ia. 

Although SNe Ia are  energetically capable of driving an outflow, the above-discussed characteristics of the gas indicate the absence of a global galactic-scale outflow. A likely explanation of the absence of an outflow and the large gas mass may be that NGC4697 is located in a galaxy group. Indeed, the temperature profile of the hot X-ray gas demonstrated the presence of a hotter component beyond the $\sim$$90\arcsec$ radius, which presumably is associated with the group emission. The hot X-ray gas inside and outside the optical extent is in pressure equilibrium, thereby suggesting that it is in hydrostatic equilibrium and not in an outflow state. This result is in good agreement with those obtained by \citet{david06}, who also concluded that the ambient pressure in a dense environment may be able to suppress the evolution of an outflow.

\subsection{Galactic-scale outflows in NGC821 and NGC3379}
Although no hot gas is detected from NGC821 and NGC3379, this does not imply the absence of X-ray gas and galactic-scale outflows. As discussed above, evolved stars eject stellar yields at a constant rate, which is proportional with the K-band luminosity (or stellar mass) of the galaxy. Based on the K-band luminosity of NGC821 and NGC3379 and using \citet{knapp92}, we estimate mass input rates of $0.17 \ \rm{M_{\odot} \ yr^{-1}}$ and  $0.13 \ \rm{M_{\odot} \ yr^{-1}}$, respectively. Hence, on a timescale of $5$ Gyr, which is comparable with the stellar age of these galaxies \citep{sanchez06}, the accumulated gas mass should be on the order of $\sim$$10^{9} \ \rm{M_{\odot}}$. Such large gas masses would imply a gas mass fraction of $\sim$$1\%$, similar to those observed in massive, X-ray luminous early-type galaxies \citep{fukazawa06}. However, such a large gas mass is not observed in NGC821 and NGC3379, implying that the X-ray gas is driven  from the galaxies. 

A plausible way to remove the X-ray gas from the potential well of the sample galaxies is via galactic-scale outflows. To derive whether SNe Ia are energetically capable of driving an outflow, we compute the available energy input from SNe Ia. Given the K-band luminosities of NGC821 and NGC3379 (Table \ref{tab:list1}) and the average SN Ia frequencies  \citep{mannucci05},  we obtain SN Ia rates of  $2.9\times10^{-3} \ \rm{yr^{-1}}$ and $2.1\times10^{-3} \ \rm{yr^{-1}}$, respectively. Assuming that each SN Ia contributes with $E_{\rm{SNIa}} = 10^{51} $ erg, we derive that at most $9.2\times10^{40} \ \rm{erg \ s^{-1}}$ and $6.7\times10^{40} \ \rm{erg \ s^{-1}}$ is available  in NGC821 and NGC3379, respectively. We confront this value with that required to drive the outflow. Therefore, we use Equation (2), the stellar mass input rates, and the velocity dispersions of NGC821 and NGC3379, which are $200.2 \ \rm{km \ s^{-1}}$ and $209.2 \ \rm{km \ s^{-1}}$, respectively. Hence, to drive an outflow from NGC821 and NGC3379, $P_{\rm{lift}}=3.1\times10^{40} \ \rm{erg \ s^{-1}}$ and $P_{\rm{lift}}=2.6\times10^{40} \ \rm{erg \ s^{-1}}$ are required. Thus, the required power is $\lesssim40\%$ of that available from SNe Ia, implying that SNe Ia are capable of driving  galactic-scale outflows in both galaxies.  

The non-detection of X-ray gas in NGC821 and NGC3379, may be explained by the low gas densities. Since the emission measure is proportional to the density square of the gas, the density plays a crucial role in determining the observed X-ray luminosity of the gas. For example, if the gas density in NGC4278 was a factor of three lower, then the X-ray luminosity would decrease by factor of nine. However, such a low luminosity gaseous emission could not be detected, therefore NGC4278 would appear as a gas-free galaxy. Following this line of argument, it is likely that both NGC821 and NGC3379 host a small amount of outflowing X-ray gas, which remains unidentified due to its relatively low density and hence low X-ray luminosity. These considerations are in good agreement with the hydrodynamical simulations of \citet{pellegrini07}, who showed that the luminosity of an outflow from NGC821 may fall below the detection  limit.

Although the radial profiles, the X-ray spectra, and the $L_X/L_K$ ratios argue for the non-detection of hot X-ray gas in NGC821 and NGC3379, we place upper limits on the gas content of these galaxies. Since the $L_X/L_K$ ratios of NGC821 and NGC3379 are in fairly good agreement with that obtained for the truly gas-free compact elliptical galaxy, M32, we estimate that the hot X-ray gas is unlikely to be responsible for more than $20\%$ of the observed thermal component in NGC821 and NGC3379. To this end, we extract the X-ray energy spectra of NGC821 and NGC3379 for their $D_{\rm{25}}$ ellipses, and describe it with a two component model, consisting of a thermal plasma emission model (APEC) and a power law component. To fit the spectra, we assumed a $0.1$ Solar abundance and Galactic column density for both galaxies. The best-fit temperatures of the thermal models are $0.62^{+0.18}_{-0.14}$ keV for NGC821 and $0.67^{+0.07}_{-0.09}$ keV for NGC3379. Assuming that $20\%$ of the thermal emission originates from hot ionized gas and assuming uniform gas density, the upper limits on the confined gas masses are $\sim$$3 \times 10^{7} \ \rm{M_{\odot}}$ for NGC821 and $\sim$$1 \times 10^{7} \ \rm{M_{\odot}}$ for NGC3379.

\subsection{Metallicity of the hot X-ray gas}
\label{sec:metal}
SNe Ia do not only  deposit energy into the X-ray emitting gas, but also enrich it with iron-peak elements. On average, each SN Ia produces $0.7 \ \rm{M_{\odot}}$ iron, thereby enriching the interstellar medium. Although the abundance of the stellar ejecta is likely to be sub-solar in low-mass ellipticals \citep{gallazzi06}, if the stellar yields undergo complete mixing with the iron-peak elements, then the abundances of the hot gas are not expected to agree with the stellar yields, but should be highly super-solar \citep{david06,bogdan08}. However, the measured abundances are $0.12^{+0.11}_{-0.03}$  in NGC4278 and  $0.14^{+0.06}_{-0.04}$ in NGC4697 (Section \ref{sec:spectra}). That is, both galaxies exhibit strongly sub-solar abundances, in conflict with the theoretical expectations. 

The observed low abundances may partly be caused by the modest energy resolution of \textit{Chandra} CCDs and the relatively low number of detected photons, which may result in large systematic uncertainties in the abundance determination. In principle, the problem of inaccurate spectral modeling can be circumvented with the application of grating spectroscopy. Although distant  galaxies with very low X-ray gas content  are not suitable for such an investigation due to the low number of  photons, studies of nearby and/or gas-rich galaxies found that abundances observed from CCD spectra are in fairly good agreement with those observed by grating spectroscopy \citep{liu10,ji09}. Therefore, it seems unlikely that the approximately order-of-magnitude difference between observed and expected abundances  originates from systematic uncertainties. 

The predicted super-solar abundances invoke the complete mixing of iron with the stellar yields. However, numerical simulations point out that in the case of an outflow, the iron is not well mixed with the ambient gas, resulting in sub-solar abundances \citep{tang09}. Therefore, the low abundances observed in NGC4278 (and possibly in NGC4697) may imply that the observed sub-solar abundances are (at least partly) due to the incomplete mixing of stellar yields and iron from SNe Ia.

In general, the low abundance in low-mass ellipticals may be due to the shallow gravitational potential well of these galaxies. At early epochs of galaxy formation, core collapse supernovae produce $\alpha$-elements. Due to supernova driven winds, a large fraction of these elements could be removed from galaxies with shallow potential wells, but are retained in galaxies with more extended halos \citep{tremonti04,gallazzi06}. At later epochs, SNe Ia contribute with iron peak elements, thereby enriching the interstellar medium. In general, low-mass ellipticals are more sensitive to AGNs and SNe Ia driven outflows at all times, which may have a major influence on their metal enrichment histories.

It is also possible that the observed low abundances are in part caused by the iron bias, which appears if a multi-temperature plasma or a temperature gradient is described with a single-temperature model \citep{buote00,baldi06}. To illustrate the importance of the iron bias, we produced a set of two-temperature thermal emission model (\textsc{APEC}) spectra for NGC4697, which we fit with a single-temperature model. The temperatures of the model spectra  were $kT_{\rm{1}}=0.2$ keV and $kT_{\rm{2}}=0.4$ keV, whereas the applied set of abundances varied from $0.2$ to $5$ Solar \citep{grevesse98}. The normalizations of the two thermal components were set, and the column density was fixed at the Galactic value. When describing the two-temperature model with a single temperature spectrum, the best-fit temperature --  independent of the abundance -- is $kT\approx0.3$ keV. However, the best-fit abundance is significantly lower than the originally assumed value, and it is in all cases subsolar. In particular, if the original abundances are set to 5/1/0.5 Solar, then the best-fit values are $\approx$0.55/0.35/0.25 Solar. Although all these values exceed that obtained for NGC4697, this examination illustrates that describing a multi-temperature plasma with an overly simplistic spectral model may have major effects on the observed abundances. We stress that this conclusion does not only refer to NGC4697 but is representative for other galaxies with similar parameters as well.  

Finally, we also mention the existence of a strong degeneracy between the emission measure and the abundance of the hot gas \citep[e.g.][]{david06}.  This degeneracy may have important consequences on the observed parameters of the hot X-ray gas. If, for example, the abundance of the gas is about Solar ($\sim$$10$ times higher than the best-fit values for NGC4278 and NGC4697), then the derived the densities, gas masses, replenishment times, and cooling times must be reduced by a factor of $\approx$$3.3$.

\section{Conclusions}
We studied the unresolved  X-ray emission, the hot ionized gas content, and the physical state of the hot X-ray gas in four low-mass elliptical galaxies (NGC821, NGC3379, NGC4278, NGC4697) based on archival \textit{Chandra} observations. Our results can be summarized as follows. \\

1. We find that the bulk of the unresolved emission in NGC821 and NGC3379 originates from  a large number of faint compact and stellar  objects, such as CVs and ABs. Despite the non-detection of hot gas, low density, and hence low luminosity X-ray gas components may be present, which, driven by the energy input of SNe Ia, outflow from the galaxies at a steady rate. 

2. We detect hot X-ray gas with a temperature of $kT\sim0.35$ keV in the central $\sim$$50\arcsec$ ($\sim$$3.9$ kpc) region of NGC4278. The total gas mass within this region is $\sim$$3\times 10^7 \ \rm{M_{\odot}}$, hence the corresponding gas mass fraction of $\sim$$0.06\%$. The X-ray gas exhibits a bipolar morphology in the northeast-southwest direction, suggesting that it is outflowing from the galaxy. We conclude that the outflowing mass can be replenished by the stellar yields of evolved stars and the outflow can be driven by the energy input of SNe Ia. Based on the existence of an outflow in NGC4278 and the energy budget of the galaxy, we place a lower limit of $\sim$$50\%$ on the supernova efficiency parameter. 

3. We show that NGC4697 hosts X-ray gas at all radii, whose temperature within a region with a $\sim$$90\arcsec$ radius is $kT\sim0.3$ keV, beyond which radius the gas temperature increases to $\sim$$0.7$ keV. We identify the hotter gas with the group atmosphere surrounding NGC4697.  Although the  X-ray gas has a significantly broader distribution than the stellar light, it does not show any asymmetries. The gas mass within the central $\sim$$90\arcsec$ region is $\sim$$2.1\times 10^8 \ \rm{M_{\odot}}$, implying a gas mass fraction of $\sim$$0.4\%$, which is comparable with those observed in luminous massive early-type galaxies. Taken together, this evidence indicates  that the X-ray gas is most likely in hydrostatic equilibrium, however a subsonic outflow cannot be excluded.

\bigskip
\begin{small}
\noindent
\'AB thanks Alexey Vikhlinin, Marat Gilfanov, and Junfeng Wang for helpful discussions. This research has made use of \textit{Chandra}  data provided by the Chandra X-ray Center. The publication makes use of software provided by the Chandra X-ray Center (CXC) in the application package CIAO. This publication makes use of data products from the Two Micron All Sky Survey, which is a joint project of the University of Massachusetts and the Infrared Processing and Analysis Center/California Institute of Technology, funded by the National Aeronautics and Space Administration and the National Science Foundation.  In this work the NASA/IPAC Extragalactic Database (NED) has been used. The authors acknowledge the use of the HyperLeda database (http://leda.univ-lyon1.fr). \'AB acknowledges support provided by NASA through Einstein Postdoctoral Fellowship grant number PF1-120081 awarded by the Chandra X-ray Center, which is operated by the Smithsonian Astrophysical Observatory for NASA under contract NAS8-03060. WF and CJ acknowledge support from the Smithsonian Institution. 
\end{small}


\begin{thebibliography}{}
\bibitem[\protect\citeauthoryear{Baldi et al.}
 {Baldi et al.}{2006}]{baldi06}
Baldi, A., Raymond, J. C., Fabbiano, G., et al., 2006, ApJ, 162, 113
\bibitem[\protect\citeauthoryear{Bell et al.}
 {Bell et al.}{2003}]{bell03}	
Bell, E. F., McIntosh, D. H., Katz, N. \& Weinberg, M. D., 2003, ApJS, 149, 289
\bibitem[\protect\citeauthoryear{Bogd\'an \& Gilfanov}
 {Bogd\'an \& Gilfanov}{2008}]{bogdan08}
  Bogd\'an, \'A., Gilfanov, M., 2008, MNRAS, 388, 56
\bibitem[\protect\citeauthoryear{Bogd\'an \& Gilfanov}
 {Bogd\'an \& Gilfanov}{2011}]{bogdan11a}
  Bogd\'an, \'A., Gilfanov, M., 2011, MNRAS, 418, 1901
\bibitem[\protect\citeauthoryear{Bogd\'an et al.}
 {Bogd\'an et al.}{2011}]{bogdan11b}
  Bogd\'an, \'A., Kraft, R. P., Forman, W. R., et al., 2011, ApJ, 743, 59
\bibitem[\protect\citeauthoryear{Buote et al.}
 {Buote et al.}{2000}]{buote00}
Buote, D. A., 2000, MNRAS, 311, 176
\bibitem[\protect\citeauthoryear{Cappellari et al.}
 {Cappellari et al.}{2011}]{cappellari11}
Cappellari, M., Emsellem, E., Krajnovi\'c, D., et al., 2011, MNRAS, 413, 813
\bibitem[\protect\citeauthoryear{David et al.}
 {David et al.}{1991}]{david91}
David, L. P., Forman, W. \& Jones, C., 1991, ApJ, 380, 39
\bibitem[\protect\citeauthoryear{David et al.}
 {David et al.}{2005}]{david05}
David, L. P., Jones, C., Forman, W. \& Murray, S. S., 2005, ApJ, 635, 1053
\bibitem[\protect\citeauthoryear{David et al.}
 {David et al.}{2006}]{david06}	
David, L. P., Jones, C., Forman, W., Vargas, I. M. \& Nulsen, P., 2006, ApJ, 653, 207
\bibitem[\protect\citeauthoryear{de Vaucouleurs et al.}
 {de Vaucouleurs et al.}{1991}]{devaucouleurs91}	
de Vaucouleurs, G., et al., Third Reference Catalogue of Bright Galaxies. Springer-Verlag. (1991)
\bibitem[\protect\citeauthoryear{Dickey \& Lockman}
 {Dickey \& Lockman}{1990}]{dickey90}
  Dickey, J. M. \& Lockman, F. J., 1990, ARA\&A, 28, 215
\bibitem[\protect\citeauthoryear{Forman et al.}
 {Forman et al.}{2007}]{forman07}
Forman, W., Jones, C., Churazov, E., et al., 2007, ApJ, 665, 1057
\bibitem[\protect\citeauthoryear{Forman et al.}
 {Forman et al.}{1985}]{forman85}
Forman, W., Jones, C. \& Tucker, W., 1985, ApJ, 293, 102
\bibitem[\protect\citeauthoryear{Fukazawa et al.}
 {Fukazawa et al.}{2006}]{fukazawa06}
Fukazawa, Y., Botoya-Nonesa, J. G., Pu, J., Ohto, A. \& Kawano, N., 2006, ApJ, 636, 698
\bibitem[\protect\citeauthoryear{Gallazzi et al.}
 {Gallazzi et al.}{2006}]{gallazzi06}	
Gallazzi, A., Charlot, S., Brinchmann, J. \& White, S. D. M., 2006, MNRAS, 370, 1106
\bibitem[\protect\citeauthoryear{Gilfanov}
 {Gilfanov}{2004}]{gilfanov04}
 Gilfanov, M., 2004, MNRAS, 349, 146 
\bibitem[\protect\citeauthoryear{Giroletti et al.}
 {Giroletti et al.}{2005}]{giroletti05}
Giroletti, M., Taylor, G. B. \& Giovannini, G., 2005, ApJ, 622, 178
\bibitem[\protect\citeauthoryear{Grevesse \& Sauval}
 {Grevesse \& Sauval}{1998}]{grevesse98}
Grevesse, N. \& Sauval, A. J., 1998, SSRv, 85, 161
\bibitem[\protect\citeauthoryear{G\"ultekin et al.}
 {G\"ultekin et al.}{2009}]{gultekin09}
G\"ultekin, K., Richstone, D. O., Gebhardt, K., et al., 2009, ApJ, 698, 198
\bibitem[\protect\citeauthoryear{Irwin et al.}
 {Irwin et al.}{2003}]{irwin03}
Irwin, J. A., Athey, A. E. \& Bregman, J. N., 2003, ApJ, 587, 356
\bibitem[\protect\citeauthoryear{Jarrett et al.}
 {Jarrett et al.}{2003}]{jarrett03}
 Jarrett, T. H., Chester, T., Cutri, R., et al., 2003, AJ, 125, 525
\bibitem[\protect\citeauthoryear{Jensen et al.}
 {Jensen et al.}{2003}]{m105distance}
  Jensen, J. B., Tonry, J. L., Barris, B. J., et al. 2003, ApJ, 583, 712
\bibitem[\protect\citeauthoryear{Ji et al.}
 {Ji et al.}{2009}]{ji09}
Ji, J., Irwin, J. A., Athey, A., Bregman, J. N. \& Lloyd-Davies, E. J., 2009, ApJ, 696, 2252
\bibitem[\protect\citeauthoryear{Kim et al.}
 {Kim et al.}{2006}]{kim06}
Kim, D.-W., Fabbiano, G., Kalogera, V., et al., 2006, ApJ, 652, 1090
\bibitem[\protect\citeauthoryear{Knapp et al.}
 {Knapp et al.}{1992}]{knapp92}
Knapp, G. R., Gunn, J. E. \& Wynn-Williams, C. G., 1992, ApJ, 399, 76
\bibitem[\protect\citeauthoryear{Kraft et al.}
 {Kraft et al.}{2011}]{kraft11}
Kraft, R. P., Forman, W. R., Jones, C., et al., 2011, ApJ, 727, 41
\bibitem[\protect\citeauthoryear{Li et al.}
 {Li et al.}{2011}]{li11}	
Li, Z., Jones, C., Forman, W. R., et al., 2011, ApJ, 730, 84
\bibitem[\protect\citeauthoryear{Li \& Wang}
 {Li \& Wang}{2007}]{li07}
Li, Z. \& Wang, Q. D., 2007, ApJ, 668, L39
\bibitem[\protect\citeauthoryear{Liu et al.}
 {Liu et al.}{2010}]{liu10}
Liu, J, Wang, Q. D., Li, Z., \& Peterson, J. R., 2010, MNRAS, 404, 1879
\bibitem[\protect\citeauthoryear{Mannucci et al.}
 {Mannucci et al.}{2005}]{mannucci05}
Mannucci, F., Della Valle, M., Panagia, N., et al., 2005, A\&A, 433, 807
\bibitem[\protect\citeauthoryear{Mathews \& Brighenti}
 {Mathews \& Brighenti}{2003}]{mathews03}
 Mathews, W. G., Brighenti, F. 2003, ARA\&A, 41, 191
\bibitem[\protect\citeauthoryear{Paturel et al.}
 {Paturel et al.}{2003}]{paturel03}
Paturel, G., Petit, C., Prugniel, P., et al.,2003, A\&A, 412, 45 
\bibitem[\protect\citeauthoryear{Pellegrini et al.}
 {Pellegrini et al.}{2007}]{pellegrini07}
Pellegrini, S., Baldi, A., Kim, et al., \& Elvis, M., 2007, ApJ, 667, 731
\bibitem[\protect\citeauthoryear{Pellegrini et al.}
 {Pellegrini et al.}{2012}]{pellegrini12}
Pellegrini, S., Wang, J., Fabbiano, G., et al., arXiv:1206.2533
\bibitem[\protect\citeauthoryear{Randall et al.}
 {Randall et al.}{2006}]{randall06}
Randall, S. W., Sarazin, C. L. \& Irwin, J. A., 2006, ApJ, 636, 200
\bibitem[\protect\citeauthoryear{Revnivtsev et al.}
 {Revnivtsev et al.}{2008}]{revnivtsev08}	
Revnivtsev, M., Churazov, E., Sazonov, S., Forman, W. \& Jones, C., 2008, A\&A, 490, 37
\bibitem[\protect\citeauthoryear{Revnivtsev et al.}
 {Revnivtsev et al.}{2009}]{revnivtsev09}	
Revnivtsev, M., Sazonov, S., Churazov, E., et al., 2009, Nature, 458, 1142
\bibitem[\protect\citeauthoryear{Revnivtsev et al.}
 {Revnivtsev et al.}{2006}]{revnivtsev06}
Revnivtsev, M., Sazonov, S., Gilfanov, M., Churazov, E. \& Sunyaev, R., 2006, A\&A, 452, 169, 178
\bibitem[\protect\citeauthoryear{S\'anchez-Bl\'azquez et al.}
 {S\'anchez-Bl\'azquez et al.}{2006}]{sanchez06}	
S\'anchez-Bl\'azquez, P., Gorgas, J., Cardiel, N. \& Gonz\'alez, J. J., 2006, A\&A, 457, 809
\bibitem[\protect\citeauthoryear{Sarazin et al.}
 {Sarazin et al.}{2001}]{sarazin01}	
Sarazin, C. L., Irwin, J. A. \& Bregman, J. N., 2001, ApJ, 556, 533
\bibitem[\protect\citeauthoryear{Sazonov et al.}
 {Sazonov et al.}{2006}]{sazonov06}
 Sazonov, S., Revnivtsev, M., Gilfanov, M., Churazov, E. \& Sunyaev, R., 2006, A\&A, 450, 117
\bibitem[\protect\citeauthoryear{Tang et al.}
 {Tang et al.}{2009}]{tang09}
Tang, S., Wang, Q. D., Mac Low, M.-M. \& Joung, M. R., 2009, MNRAS, 398, 1468
\bibitem[\protect\citeauthoryear{Thornton et al.}
 {Thornton et al.}{1998}]{thornton98}
Thornton, K., Gaudlitz, M., Janka, H.-Th. \& Steinmetz, M., 1998, ApJ, 500, 95
\bibitem[\protect\citeauthoryear{Tonry et al.}
 {Tonry et al.}{2001}]{tonry01}
 Tonry, J. L., Dressler, A., Blakeslee, J. P., et al., C. B., 2001, ApJ, 546, 681
\bibitem[\protect\citeauthoryear{Tremonti et al.}
 {Tremonti et al.}{2004}]{tremonti04}	
Tremonti, C. A., Heckman, T. M.; Kauffmann, G., et al., 2004, ApJ, 613, 898
\bibitem[\protect\citeauthoryear{Trinchieri et al.}
 {Trinchieri et al.}{2008}]{trinchieri08}	
Trinchieri, G., Pellegrini, S., Fabbiano, G., et al., 2008, ApJ, 688, 1000
\end{thebibliography}
\end{document}